\documentclass[sigconf]{acmart}

\usepackage{pdfpages}
\usepackage{amsfonts}
\usepackage{pifont}
\usepackage{colortbl}
\usepackage{xcolor}
\usepackage{microtype}
\usepackage{multirow, booktabs}
\usepackage{adjustbox}
\usepackage{enumitem}
\usepackage{float}
\usepackage{bm}
\usepackage{bbm}
\usepackage{wrapfig,lipsum,booktabs}

\definecolor{purple}{rgb}{0.5,0,1}
\definecolor{dcyan}{rgb}{0.2,0.6,0.5}
\definecolor{darkgreen}{rgb}{0,200,0}
\definecolor{darkorange}{rgb}{138, 50, 0}
\definecolor{light-gray}{gray}{0.95}
\definecolor{darkgreen}{RGB}{0,140,0}
\definecolor{darkred}{RGB}{200,0,0}
\definecolor{lightgreen}{RGB}{231,255,219}
\definecolor{lightred}{RGB}{252,231,234}
\definecolor{lightyellow}{RGB}{250,253,191}
\definecolor{DarkRed}{RGB}{130,25,0}

\AtBeginDocument{
  \providecommand\BibTeX{{
    \normalfont B\kern-0.5em{\scshape i\kern-0.25em b}\kern-0.8em\TeX}}}

\copyrightyear{2022}
\acmYear{2022}
\setcopyright{acmlicensed}
\acmDOI{10.1145/3523227.3546767}

\acmConference[RecSys '22]{Sixteenth ACM Conference on Recommender Systems}{September 18--23, 2022}{Seattle, WA, USA}
\acmBooktitle{Sixteenth ACM Conference on Recommender Systems (RecSys '22), September 18--23, 2022, Seattle, WA, USA}
\acmPrice{15.00}
\acmISBN{978-1-4503-9278-5/22/09}

\begin{document}

\title[Recommendation as Language Processing: P5]{Recommendation as Language Processing (RLP): \\A Unified Pretrain, Personalized Prompt \& Predict Paradigm (P5)}

\author{Shijie Geng, Shuchang Liu, Zuohui Fu, Yingqiang Ge, Yongfeng Zhang}
\affiliation{
 \institution{Department of Computer Science, Rutgers University, NJ 08854
 \country{US}
 }}
\email{{sg1309, shuchang.syt.liu, zuohui.fu, yingqiang.ge, yongfeng.zhang}@rutgers.edu}

\begin{abstract}
For a long time, different recommendation tasks typically require designing task-specific architectures and training objectives. As a result, it is hard to transfer the learned knowledge and representations from one task to another, thus restricting the generalization ability of existing recommendation approaches, e.g., a sequential recommendation model can hardly be applied or transferred to a review generation method. To deal with such issues, considering that language can describe almost anything and language grounding is a powerful medium to represent various problems or tasks, we present a flexible and unified text-to-text paradigm called ``Pretrain, Personalized Prompt, and Predict Paradigm'' (\textbf{P5}) for recommendation, which unifies various recommendation tasks in a shared framework. In P5, all data such as user-item interactions, user descriptions, item metadata, and user reviews are converted to a common format --- natural language sequences. The rich information from natural language assists P5 to capture deeper semantics for personalization and recommendation. Specifically, P5 learns different tasks with the same language modeling objective during pretraining. Thus, it serves as the foundation model for various downstream recommendation tasks, allows easy integration with other modalities, and enables instruction-based recommendation based on prompts. P5 advances recommender systems from shallow model to deep model to large model, and will revolutionize the technical form of recommender systems towards universal recommendation engine. With adaptive personalized prompt for different users, P5 is able to make predictions in a zero-shot or few-shot manner and largely reduces the necessity for extensive fine-tuning. On several recommendation benchmarks, we conduct experiments to show the effectiveness of P5. To help advance future research on Recommendation as Language Processing (RLP), Personalized Foundation Models (PFM), and Universal Recommendation Engine (URE), we release the source code, dataset, prompts, and pretrained P5 model at \url{https://github.com/jeykigung/P5}. Meanwhile, P5 is also hosted on Hugging Face at \url{https://huggingface.co/makitanikaze/P5}.
\end{abstract}


\maketitle

\section{Introduction}
For the past decades, recommender systems have witnessed significant advancements and played an essential role in people's daily life, helping their micro decisions and fulfilling their demands with outstanding accuracy. In retrospect, we can summarize the development trend of modern recommender systems -- towards a more comprehensive system that accommodates diverse features and a wide spectrum of application scenarios.

On one hand, feature engineering and learning in recommender systems has evolved greatly from simple to complex. 
In early ages, recommender systems typically adopt logistic regression or collaborative filtering~\cite{resnick1994grouplens,sarwar2001item,linden2003amazon,koren2009matrix} which utilize user-item interaction records to model users' behavioral patterns.
Later on, the contextual features such as user profile and item metadata are further integrated into the system through more sophisticated models such as factorization machines~\cite{rendle2010factorization} and GBDT~\cite{he2014practical}. 
Recently, deep neural network models~\cite{cheng2016widedeep,guo2017deepfm,chen2021neural,zhang2017joint} facilitate crossing and combination among even more diverse and sophisticated features.
As a result, these models gain better representation ability compared with traditional feature engineering based approaches.

\begin{figure*}[t!]
\centering
\includegraphics[width=15.0cm]{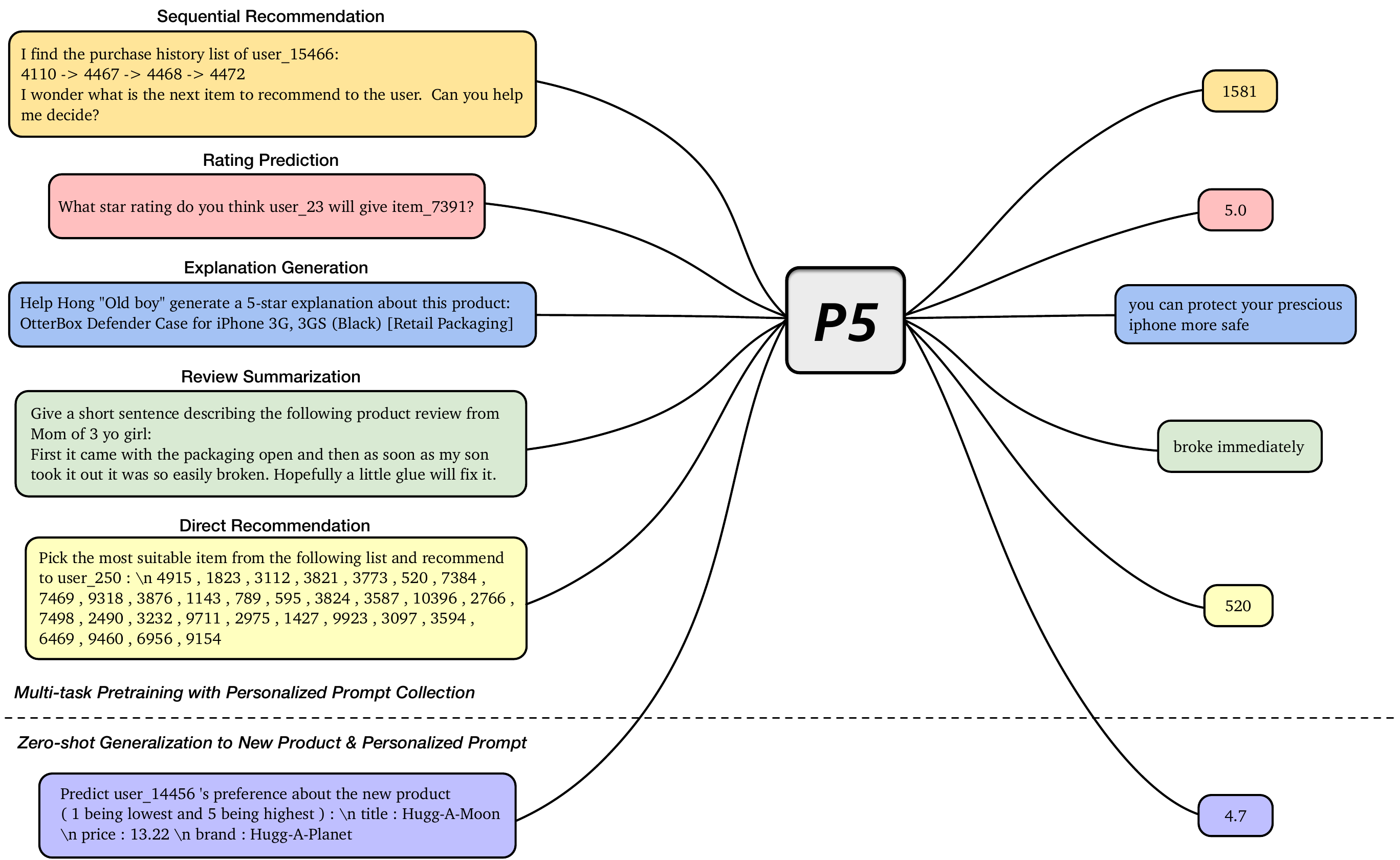}
\caption{P5 pretrains on an encoder--decoder Transformer model that takes in textual inputs and produces target responses. We trained P5 on a multitask collection of personalized prompts. After multitask prompt-based pretraining on recommendation datasets, P5 achieves the capability of zero-shot generalization to unseen personalized prompts and new items.}
\vspace{-10pt}
\label{fig:teaser}
\end{figure*}

On the other hand, more recommendation tasks have emerged. 
Except for classical rating prediction and direct user-item matching-based recommendation tasks, recent works are broadening the spectrum to new tasks and scenarios such as sequential recommendation \cite{sun2019bert4rec,zhou2020s3,hidasi2016gru4rec,tang2018personalized}, conversational recommendation \cite{zhang2018towards,christakopoulou2016towards,sun2018conversational}, explainable recommendation \cite{zhang2014explicit,zhang2020explainable,li2021personalized,tan2021counterfactual,xian2019reinforcement,ge2022survey} and so on.
While the approaches to the aforementioned recommendation tasks are often proposed separately, there is an evident trend of utilizing multiple recommendation tasks to jointly learn the transferable representations~\cite{shin2021one4all,shin2021scaling,yuan2021one,li2021personalized}.
Although existing recommender systems achieved great success, there is still a considerable gap between current solutions and the foreseeable intersection of the aforementioned trends -- a comprehensive recommender system that can accommodate diverse features and different types of tasks. Since recommendation tasks usually share a common user--item pool and have overlapping contextual features, we believe it is promising to merge even more recommendation tasks into a unified framework so that they can implicitly transfer knowledge to benefit each other and enable generalization to other unseen tasks.

Inspired by the recent progress in multitask prompt-based training~\cite{aribandi2022ext5, sanh2022multitask, wei2022finetuned}, in this work, we propose a unified ``Pretrain, Personalized Prompt \& Predict Paradigm'' (denoted as \textbf{P5}).
We show that P5 is possible to learn multiple recommendation related tasks together through a unified sequence-to-sequence framework by formulating these problems as prompt-based natural language tasks, where user--item information and corresponding features are integrated with personalized prompt templates as model inputs.
P5 sheds light on a promising technical route for unified and instruction-based recommendation. It has \emph{three} main advantages:

\vspace{0.05cm}
\noindent 1) P5 deeply immerses recommendation models into a full language environment, where all recommendation tasks are reformulated to NLP tasks with the help of personalized prompts.
Since language grounding is sufficiently flexible and powerful to express various kinds of features in text templates, so there is no need to design feature-specific encoders. As a result, P5 can exploit the abundant semantics and knowledge inside the training corpora;

\vspace{0.05cm}
\noindent 2) P5 integrates multiple recommendation tasks into a shared text-to-text encoder-decoder architecture and trains them with the same language modeling loss rather than designing task-specific architectures and objective functions. In other words, P5 treats all personalized tasks as a conditional text generation problem;

\vspace{0.05cm}
\noindent 3) Trained with instruction-based prompts, P5 attains sufficient zero-shot performance when generalizing to novel personalized prompts or unseen items in other domains. 

In our experiments, we study how P5 performs compared with task-specific approaches on all five task families as well as evaluating P5's zero-shot generalization ability. We also conduct several ablation studies to justify the design details of P5 framework. Overall, our main contributions can be outlined as follows:
\vspace*{-1pt}
\begin{itemize}[leftmargin=*,noitemsep]
    \item To the best of our knowledge, this is the first work to propose a unified ``Pretrain, Personalized Prompt \& Predict Paradigm'' which integrates various recommendation related tasks into a shared conditional language generation framework.
    \item We create a collection of personalized prompts that cover five different recommendation task families.
    \item According to the experimental results, P5 achieves promising performances on the five task families when taking seen prompt templates as model inputs. 
    \item P5 shows sufficient zero-shot generalization ability for novel personalized prompts and new items in unseen domains.
\end{itemize}

\begin{figure*}[t!]
\centering
\includegraphics[width=0.95\linewidth]{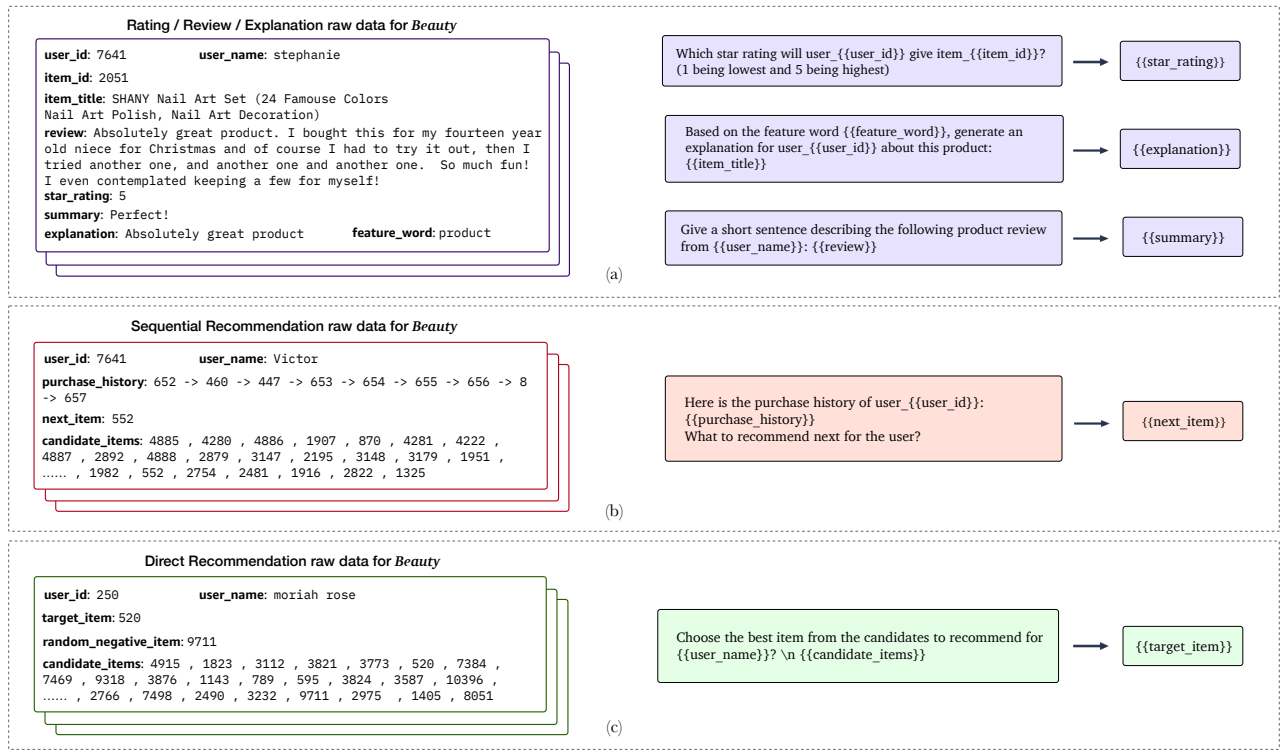}
\caption{Building input--target pairs from raw data according to our designed personalized prompt templates -- simply substituting the fields in the prompts with the corresponding information in raw data. The raw data for the five task families of P5 are from three separate sources. Specifically, rating/review/explanation prompts (a) have shared raw data. Sequential recommendation (b) and direct recommendation (c) uses similar raw data, but the former particularly requires the user interaction history. The complete collection of P5 personalized prompts are provided in the Appendix.}
\vspace{-10pt}
\label{fig:templates}
\end{figure*}

\section{Related Work}
\noindent\textbf{Unified Frameworks.~}
Many prior works have pursued to solve various tasks in a unified model. As early pioneers, T5~\cite{2020t5} and GPT-3~\cite{brown2020language} unifies NLP downstream tasks through text-to-text encoder--decoder framework and autoregressive language modeling, respectively. They both allow effective knowledge sharing among different tasks based on a common pretrained language model. 
Following this trend, recent advances started to focus on unifying large-scale language tasks~\cite{sanh2022multitask, aribandi2022ext5,wei2022finetuned} or cross-modality applications~\cite{yang2021crossing, cho2021vlt5, wang2022unifying} through a shared sequence-to-sequence framework, where different types of tasks and modalities are all expressed in the format of natural language. However, aforementioned methods never consider personalization in their sequence-to-sequence models.
Recently, a line of work~\cite{shin2021one4all,shin2021scaling,yuan2021one} attempt to learn universal user representations which are easily transferrable to downstream tasks. One limitation of these methods is that they still require additional finetuning on downstream datasets. In contrast, our P5 first takes personalization into an encoder-decoder Transformer model that can generalize to a wide spectrum of recommendation related application scenarios -- tasks that naturally require personalization. Moreover, with the help of prompt-based pretraining, P5 acquires zero-shot generalization ability when transferring to unseen prompts and items.

\vspace{0.1cm}
\noindent\textbf{Prompt Learning.~}
The success of GPT series especially GPT-3~\cite{brown2020language} marked the beginning of prompt's popularization on NLP tasks. Trained with huge language data from the Web, GPT-3 exhibited the capability of solving NLP tasks when provided a number of input-output examples as exemplar prompts. Besides exemplar prompts, many prompt design methods have proliferated following the ``pre-train, prompt, and predict'' paradigm~\cite{liu2021pre}. 
One type of the methods~\cite{jiang2020can, shin2020autoprompt, liu2021makes, gao2021making, lu2022fantastically} explored prompt search for proper discrete prompts. Meanwhile, another line of work~\cite{zhou2021coop, li2021prefix, liu2021gpt, lester2021power, gu2021ppt, qin2021learning} exploited continuous vector embeddings as prompts.
Compared with the aforementioned prompt types, instruction-based prompts contain detailed task descriptions and adhere more to the natural language format. Since instruction-based prompts are flexible and close to how humans communicate with each other, several pioneer works~\cite{weller2020learning,efrat2020turking} claim that learning from crowd-sourced NLP datasets is a promising route for general purpose NLP systems. Recent works such as FLAN~\cite{wei2022finetuned} and T0~\cite{sanh2022multitask} finetuned pretrained language models on large-scale NLP datasets verbalized via human-readable prompts. As a result, such multitask prompt-based tuning brings powerful models that exhibit strong zero-shot ability on unseen tasks. Inspired by the success of these approaches, we create a collection of personalized prompts and then train a sequence-to-sequence model on a variety of recommendation related tasks verbalized according to the constructed personalized prompts.

\begin{figure*}[t!]
\centering
\includegraphics[width=\linewidth]{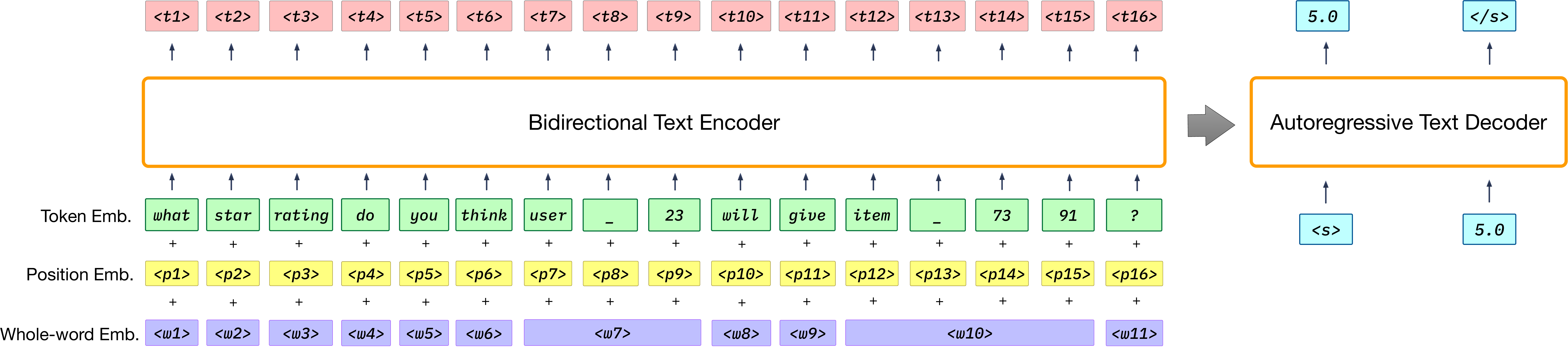}
\vspace{-10pt}
\caption{An illustration of the P5 architecture. For the example prompt input ``What star rating do you think user\_23 will give item\_7391?'', P5 adopts an encoder--decoder framework: first encodes the input with a bidirectional text encoder, and then generates the answer through a text decoder autoregressively. In contrast to task-specific recommendation models, our P5 relies on multitask prompt-based pretraining on a large-scale personalized prompt collection, which makes P5 able to adapt to different task families and even generalize to novel personalized prompts.}
\vspace{-10pt}
\label{fig:framework}
\end{figure*}

\vspace{0.1cm}
\noindent\textbf{NLP for Recommendation.~} 
Recommendation has been interacting with NLP techniques for a long time. 
The main work mostly address four lines of research: 1) explainable recommendation~\cite{zhang2014explicit,EACL17-Att2Seq,li2017neural,li2020generate,zhang2020explainable,li2021personalized,chen2019personalized} where NLP models help generating text explanations for a given recommendation; 2) sequential recommendation as language modeling~\cite{sun2019bert4rec,zhou2020s3,de2021transformers4rec} which considers user interaction histories as word token sequences; 
3) text feature extraction~\cite{zhang2017joint,zheng2017joint,wu2019neural} which aims to extract informative text encodings that can improve the performance of recommendation;
and 4) conversational recommendation~\cite{zhang2018towards,christakopoulou2016towards,sun2018conversational,fu2021popcorn,fu2021hoops,jannach2021survey,fu2020tutorial} that reasons the intent of users and gives recommendation in an interactive dialog format.
In our work, we explicitly covers the tasks of sequential recommendation and explanation generation, and additionally offers insights on how to formulate a unified NLP framework for other recommendation problems including rating prediction, top-k recommendation, and review summarization.
Furthermore, pretrained with instruction-based prompts that share similarity with conversational recommendation, our P5 benefits from the natural language environment and improves the performance on a series of recommendation tasks.

\vspace{0.1cm}
\noindent\textbf{Zero-shot and Cold Start Recommendation.~}
Recommender systems' performances heavily rely on the available training data, but there are always zero-shot cases where the history records are limited.
The evidences of performing well on such startup cases signal a good generalization ability of recommendation models.
One widely studied problem under this setting is the cold-start recommendation where users~\cite{lam2008addressing} or items~\cite{schein2002methods} are new to the system with no previous interaction records.
Solutions to this problem either learn to model content features~\cite{pazzani2007content,gantner2010learning,li2019zero,shi2019adaptive} so that inference can be made without interaction records or learn to transfer representations from auxiliary domains~\cite{singh2008relational,man2017cross,zhu2021cdrsurvey,shin2021one4all,yuan2021one}.
Another line of work for zero-shot or few-shot recommendation discusses the quick adaptation to the new domain instead of providing recommendation for cold-start cases only.
Solutions typically follow the meta learning~\cite{vartak2017meta,lee2019melu} or causal learning~\cite{li2022causal} frameworks that make the model robust to domain adaptations.
In our work, we ask P5 model pretrained on an auxiliary domain to solve tasks on target domains, where the users are known to P5 but the items have never been seen by the model before.

\section{Personalized Prompt Collection}
To facilitate the multitask prompt-based pretraining for recommendation, we create a collection of personalized prompt templates. The collection covers five different task families -- \textbf{rating}, \textbf{sequential recommendation}, \textbf{explanation}, \textbf{review}, and \textbf{direct recommendation}. Each of these task families contains multiple personalized prompts to help P5 discover various aspects about users and items. As mentioned in \cite{sanh2022multitask}, a prompt is considered as consisting of an input template and a target template, along with a collection of associated metadata. In this work, we further define a \emph{personalized prompt} as a prompt that includes personalized fields for different users and items. For example, a user's preference can be indicated through either an ID number or a description of the user such as name, gender, age, etc. Moreover, the expected model output of a given personalized prompt should also vary according to its item field. This implies the change of user's preferences towards different items. Such item fields can be represented by either item ID numbers or item metadata that contains detailed descriptions.

We designed basic P5 personalized prompt collection for each task family. 
For \textbf{rating} prediction task family, we divide the prompts into three categories: 1) Given the information about a user and an item, directly predict the rating score ranging from 1 to 5; 2) Predict whether a user will rate an item a given score. The expected output is yes or no; 3) Predict if a user likes or dislikes an item. Here we consider a star rating equal to or greater than 4 to be a \emph{like} preference of the user, whereas lower scores indicate a \emph{dislike} preference. 
For \textbf{sequential} recommendation task family, we create three types of prompts: 1) Directly predict the next item based on user interaction history; 2) Given user interaction history, choose the possible next item from a candidate list, where only one item is positive; 3) Based on user interaction history, predict whether a given item will be interacted next by the user.
For \textbf{explanation} task family, we ask P5 model to generate a textual explanation to justify a user's preference towards a given item. There are two prompt categories in this task family: 1) Directly generate an explanation sentence with user/item information; 2) Generate explanation based on a feature word as hint \cite{li2021personalized}. For each category, there could be other auxiliary information included such as the review headline and the star rating.
For \textbf{review} related task family, we create two types of prompts: 1) Summarize review comment to a shorter review title; 2) Predict the corresponding rating score based on the given review comment.
For \textbf{direct} recommendation, we also create two types of prompts: 1) Predict whether to recommend an item to a user, the answer should be yes or no; 2) Select the most suitable item from a list of candidate items to recommend to the user.
We provide some example prompts in Figure \ref{fig:templates}, and the complete collection of personalized prompts are provided in the Appendix.

With the prompts, we can directly build input--target pairs from raw data. As illustrated in Figure~\ref{fig:templates}, we can simply substitute the fields in braces with the corresponding information in the raw data and thus create training input--target pairs or zero-shot testing personalized prompts. 
The training data and pre-training tasks will distill the rich semantics from diverse modalities into the user and item tokens for preference understanding and personalization.
Note that we divide the raw data into three parts---rating/review/explanation share the same raw data, while sequential and direct recommendation differ in terms of whether to use interaction history as input information. 
During pretraining, we mix the input--target pairs from different task families together to serve as the training data. To enhance P5's robustness and zero-shot generalization, for each raw datum, we only sample a portion of rather than all of the personalized prompts in each task family. In sequential and direct recommendation task families, we also randomly select a group of negative items for those prompts that require a candidate list.

\section{The P5 Paradigm and Model}
\subsection{The P5 Architecture}
The collection of personalized prompts introduced in the previous section makes it convenient to create a large amount of available pretraining data that covers a wide range of recommendation related tasks. Thanks to the prompt templates, all pretraining data shares a unified format of input--target token sequences, which breaks the boundaries among different tasks. We claim that pretraining multiple recommendation tasks under a unified framework of conditional generation can facilitate all involving tasks together. By immersing P5 in the full language environment throughout the pretraining stage, we also expect its zero-shot generalization capability of understanding unseen personalized prompts with detailed item descriptions. That is the reason why P5 is called a unified ``Pretrain, Personalized Prompt, and Predict Paradigm''.

In terms of the model architecture, our P5 is established upon a basic encoder--decoder framework. We employ Transformer~\cite{vaswani2017attention} blocks to build both the encoder and decoder. Suppose the embeddings of an input token sequence is $\mathbf{x} = \left[x_1, \cdots, x_n\right]$. As depicted in Figure~\ref{fig:framework}, before feeding the embedding sequence into the bidirectional text encoder $\mathcal{E}(\cdot)$, we add positional encodings $\mathcal{P}$ to the raw embeddings to capture their position information in the sequence. Furthermore, to make P5 aware of the personalized information contained in the input sequence, we also apply whole-word embeddings $\mathcal{W}$ to indicate whether consecutive sub-word tokens are from the same original word. For instance, if we directly represent the item with ID number 7391 as ``item\_7391'', then the word will be split into 4 separate tokens (i.e., ``item'', ``\_'', ``73'', ``91'') by SentencePiece tokenizer~\cite{sennrich2016neural}. With the assistance of the shared whole-word embedding ``$\langle$w10$\rangle$'' (e.g., in Figure \ref{fig:framework}), P5 can better recognize the important field with personalized information.
Another alternative is to represent each user/item by an independent extra token (e.g., ``$\langle$item\_7391$\rangle$''). However, this may incur huge amounts of additional tokens when there is a large pool of users and items. Hence, in this paper, we adopt multiple sub-word units to represent a user or item.

Afterwards, the text encoder takes the sum of the aforementioned three embeddings $\mathbf{e} = \left[e_1, \cdots, e_n\right]$ and outputs their contextualized representations $\mathbf{t} = \left[t_1, \cdots, t_n\right] = \mathcal{E}(\mathbf{e})$. The decoder $\mathcal{D}(\cdot)$ then attends to both the previously generated tokens $\mathbf{y}_{<j}$ and the encoder output $\mathbf{t}$ and predicts the probability distribution of future tokens: $P_{\theta}\left(\mathbf{y}_{j} \mid \mathbf{y}_{<j}, \mathbf{x}\right) = \mathcal{D}(\mathbf{y}_{<j}, \mathbf{t})$. During the pretraining stage, P5 learns the model parameters $\theta$ by minimizing the negative log-likelihood of label
tokens $\mathbf{y}$ conditioned on input text $\mathbf{x}$ in an end-to-end manner:
\begin{equation}
\label{eq:1}
\mathcal{L}_{\theta}^{\mathrm{P5}}=-\sum_{j=1}^{|\mathbf{y}|} \log P_{\theta}\left(\mathbf{y}_{j} \mid \mathbf{y}_{<j}, \mathbf{x}\right)
\end{equation}
This same objective function is shared by all recommendation tasks under P5. As a result, we unify recommendation tasks with one model, one loss, and one data format.

\vspace{-3pt}
\subsection{Recommendation with Pretrained P5}
After pretraining, P5 can directly perform different tasks with either seen or unseen personalized prompts. For rating, explanation, and review tasks, we simply use greedy decoding to generate  answers. In contrast, sequential and direct recommendation tasks usually require an item list as target output. In view of this, for sequential recommendation, we apply beam search to generate a list of potential next items and evaluate it under the all-item setting. For direct recommendation, we predict the recommended items from a candidate set $\mathbf{S} = \{S_1, \cdots, S_m\}$, where only one of the $m$ candidates is positive. Here, we also use beam search to decode a list of potential target items with the highest scores and then conduct evaluations. Both of the above decoding processes can be written as:
\begin{equation}
        \mathbf{C} = [C_1, \cdots, C_B] = \mathrm{Beam\_Search}(\mathcal{D}, \mathbf{t}, B)
        \label{eq:2}
\end{equation}
where $B$ denotes the beam size and $\mathbf{C}$ is the output item list.

\vspace{-4pt}
\section{Experiments}
In this section, we evaluate the performance of the proposed P5 approach on real-world data and compare it with various representative methods targeting at different task families. Through the performance comparison and ablation studies, we aim to answer the following research questions regarding our unified ``Pretrain, Personalized Prompt, and Predict Pargadigm'' (P5):
\begin{itemize}[leftmargin=*,noitemsep]
\item \textbf{RQ1:} How does our unified P5 framework perform compared with task-specific methods on all five task families? 

\item \textbf{RQ2:} Does P5 have enough zero-shot generalization ability when transferring to unseen personalized prompts for either existing or new items?

\item \textbf{RQ3:} How do scaling factors such as model size, number of task families, and number of prompts affect the performance of P5? 

\item \textbf{RQ4:} Which is a better way to implement personalization in P5: adopting an independent extra token for each user or item (e.g., ``$\langle$user\_23$\rangle$'') or the default setting, i.e., tokenizing each user or item into multiple sub-word units (e.g., ``user'', ``\_'', ``23'')?

\item \textbf{RQ5:} How long does it take for P5 to conduct pretraining? Is it efficient to make inference with the pretrained P5 model? We provide statistics on training and inference time in the Appendix.
\end{itemize}

\begin{table}[t!]
\centering
\small
\caption{Basic statistics of the experimental datasets.}
\vspace{-10pt}
\begin{tabular}{lrrrrr}
\toprule
Dataset &  {\textbf{Sports}} &  {\textbf{Beauty}} & {\textbf{Toys}} & {\textbf{Yelp}}\\
\cmidrule{1-5}
\#Users    &  35,598   &  22,363   &  19,412  &  30,431\\
\#Items    &  18,357   &  12,101  & 11,924  & 20,033\\
\#Reviews  & 296,337   &  198,502   & 167,597 & 316,354\\
\#Sparsity (\%) &  0.0453  & 0.0734  &  0.0724 &  0.0519\\
\bottomrule
\end{tabular}
\vspace{-10pt}
\label{tab:stats}
\end{table}

\subsection{Experimental Setup} 
\label{sec:setup}
\noindent\textbf{Datasets.~} 
We conduct extensive experiments over \emph{four} real-world datasets. 
The Amazon\footnote{\url{https://nijianmo.github.io/amazon/}} datasets are collected from \textit{Amazon.com} platform with user ratings and reviews on 29 categories of products. In this paper, we adopt three of them to evaluate our method, namely \textit{Sports \& Outdoors}, \textit{Beauty}, as well as \textit{Toys \& Games}.
Besides, Yelp\footnote{\url{https://www.yelp.com/dataset}} dataset contains a large number of user ratings and reviews for business recommendation. We follow \cite{zhou2020s3} and use transaction records between January 1, 2019 to December 31, 2019.
Due to space limit and that the results on Yelp show similar trends with other datasets, we put the experimental results on \emph{Yelp} dataset in the Appendix.
The detailed statistics of these datasets are presented in Table~\ref{tab:stats}.

\vspace{0.07cm}
\noindent\textbf{Task splits.~} For rating, explanation, and review task families, we randomly split each dataset into training (80\%), validation (10\%) and testing (10\%) sets, and ensure that there is at least one instance included in the training set for each user and item.  
To obtain the ground-truth explanations, following the natural language explanation works \cite{li2020generate,li2021personalized}, we first extract item feature words from the reviews with the help of the Sentires toolkit\footnote{\url{https://github.com/evison/Sentires}}\cite{SIGIR14-Sentires,zhang2014explicit}, and then extract the sentences from reviews that comment on one or more item feature words as users' explanation about their preference.
In terms of sequential recommendation task family, for each user interaction sequence, the last item is used as the test data, the item before the last one is used as the validation data, and the remaining data is used for training. To avoid data leakage during pretraining, we follow the training split of sequential recommendation to build the training set for direct recommendation task family. 

\vspace{0.07cm}
\noindent\textbf{Implementation Details.~} 
Our P5 model utilizes the pretrained T5 checkpoints ~\cite{2020t5} as backbone. According to the size of T5 backbone, we create two versions of P5, namely P5-small (\textbf{P5-S}) and P5-base (\textbf{P5-B}). For P5-small, there are $6$ layers for both encoder and decoder, the model dimensionality is $512$ with $8$-headed attention, and the number of parameters is 60.75 million.
For P5-base, encoder and decoder both have $12$ Transformer blocks. The model has an embedding dimensionality of $768$ and a $12$-headed attention, and the number of parameters is 223.28 million.
For tokenization, we use the SentencePiece~\cite{sennrich2016neural} tokenizer with a vocabulary size of $\text{32,128}$ for parsing sub-word units. 
We pretrain P5 for $10$ epochs with AdamW optimization~\cite{loshchilov2018decoupled} on four NVIDIA RTX A5000 GPUs. The batch size is set to $16$ for P5-base and $32$ for P5-small. We choose $1\times 10^{-3}$ as the peak learning rate and set the maximum length of input tokens to $512$. The warmup strategy is used to adjust the learning rate during training, the warmup stage is set to be the first 5\% of all iterations. When negative sampling is needed for training, we use 1:1 positive vs. negative sampling for both P5 and baselines.

Our default \emph{pretrain--predict combination} adopts the \textbf{last} prompt in each task family for zero-shot evaluation while all remaining prompts are utilized for multitask prompted pretraining.
For rating prediction, we use Gaussian sampling to convert the original integer scores to float numbers rounded to $1$ decimal place. In this way, we can avoid overfitting the limited score types. After this change, we increase the number of score classes from $5$ to $41$.
For sequential recommendation, we set the beam size $B$ to $20$. For direct recommendation, the beam size is also $20$ and the candidate pool contains $100$ items, which consist of one ground-truth item and $99$ sampled negative ones that the user has not interacted with.

\vspace{0.05cm}
\noindent\textbf{Metrics.~} For rating prediction, we adopt Root Mean Square Error (RMSE) and Mean Absolute Error (MAE). For sequential recommendation and direct recommendation tasks, we employ top-$k$ Hit Ratio (HR@$k$) and Normalized Discounted Cumulative Gain (NDCG@$k$) to evaluate the performance and report HR@{1, 5, 10} and NGCG@{5, 10}. For explanation generation and review summarization, we evaluate different methods with BLEU-4, as well as ROUGE-1, ROUGE-2, and ROUGE-L. RMSE and MAE are ``the lower, the better'', while all other metrics are ``the higher, the better''. For all tables in the following, \textbf{bold} numbers refer to the best performance, while \underline{underlined} numbers indicate the second best performance. 

\subsection{Baselines for Multiple Tasks}
To demonstrate P5's competence on a wide range of recommendation related tasks, we gather a collection of representative approaches for difference task families. 

\vspace{0.05cm}
\noindent\textbf{Rating Prediction and Direct Recommendation.~} These tasks take the user--item rating/interaction data, but no content or side information is provided. We aim to justify whether the models are able to provide accurate rating prediction or recommendation lists that align with the user preferences. We use \textbf{MF}~\cite{koren2009matrix} and \textbf{MLP}~\cite{cheng2016widedeep} under mean square root loss as rating prediction baselines. For direct recommendation, we use \textbf{BPR-MF} \cite{rendle2009bpr}, \textbf{BPR-MLP} \cite{cheng2016widedeep}, and a state-of-the-art contrastive learning-based collaborative filtering model \textbf{SimpleX} \cite{mao2021simplex} as baselines.

\vspace{0.05cm}
\noindent\textbf{Sequential Recommendation.~} We adopt several representative sequential recommendation approaches as our baselines. \textbf{Caser}~\cite{tang2018personalized} treats sequential recommendation as a Markov Chain and employs convolutional neural networks to model user interests. \textbf{HGN}~\cite{ma2019hierarchical} adopts a hierarchical gating networks to learn user behaviors from the perspectives of both long and short terms. \textbf{GRU4Rec}~\cite{hidasi2016gru4rec} is originally proposed for session-based recommendation. It utilizes GRU~\cite{cho2014learning} to model the user click history sequence. \textbf{BERT4Rec}~\cite{sun2019bert4rec} mimics the BERT-style masked language modeling and learns a bidirectional representation for sequential recommendation. \textbf{FDSA}~\cite{zhang2019feature} focuses on the feature transition patterns by modeling feature sequence with a self-attention module. 
\textbf{SASRec}~\cite{kang2018self} adopts self-attention mechanism in a sequential recommendation model, which reconciles the properties of Markov Chains and RNN-based approaches.
\textbf{S$^3$-Rec}~\cite{zhou2020s3} leverages self-supervised objectives to help sequential recommendation model better discover the correlations among different items and their attributes. We use the implementation of S$^3$-Rec and its baselines for comparison\footnote{https://github.com/RUCAIBox/CIKM2020-S3Rec}.

\vspace{0.05cm}
\noindent\textbf{Explanation Generation.~} For performance comparison, we consider several baselines with regard to the task of explanation generation. \textbf{Attn2Seq}~\cite{EACL17-Att2Seq} learns to encode attributes into vectors, and then invokes an attention mechanism to generate reviews conditioned on the attribute vector. \textbf{NRT}~\cite{li2017neural} utilizes GRU~\cite{cho2014learning} to generate explanations based on user and item IDs.
\textbf{PETER}~\cite{li2021personalized} is a simple and effective framework that attempts to utilize user and item IDs to generate explanations. It is built upon a modified attention mask of the Transformer architecture. There is also a variant \textbf{PETER+}, which takes a hint feature word to assist the explanation generation.

\vspace{0.05cm}
\noindent\textbf{Review Related.~} For review summarization, we adopt pretrained \textbf{T0}~\cite{sanh2022multitask} and \textbf{GPT-2}~\cite{radford2019language} checkpoints hosted by Hugging Face\footnote{\url{https://huggingface.co/}} as baselines. For review preference prediction, we only use \textbf{T0} to make comparisons because GPT-2 cannot perform this task.

\begin{table}[t!]
\centering
\footnotesize
\caption{Performance comparison on rating prediction.}
\vspace{-10pt}
\begin{tabular}{ccccccc}
\toprule
\multirow{2.5}{*}{Methods} & \multicolumn{2}{c}{\textbf{Sports}} & \multicolumn{2}{c}{\textbf{Beauty}} &  \multicolumn{2}{c}{\textbf{Toys}} \\
\cmidrule(lr){2-3}\cmidrule(lr){4-5}\cmidrule(lr){6-7}
 & RMSE  & MAE & RMSE  & MAE & RMSE  & MAE \\
\cmidrule{1-7}
MF    &  \bf 1.0234  & 0.7935  & \bf 1.1973 & 0.9461 & \bf 1.0123 & 0.7984    \\
MLP  & 1.1277 & 0.7626  & 1.3078  & 0.9597  & 1.1215 & 0.8097  \\
P5-S {\color{purple}(1-6)}   &  1.0594  & \bf 0.6639  & 1.3128   & \bf 0.8428  & 1.0746  &  \underline{0.7054}   \\
P5-B {\color{purple}(1-6)}   &  1.0357  & 0.6813  & \underline{1.2843} & 0.8534 & 1.0544  & 0.7177    \\
P5-S {\color{orange}(1-10)}  &  1.0522  & \underline{0.6698}  & 1.2989   & \underline{0.8473}  & 1.0550  &  0.7173  \\
P5-B {\color{orange}(1-10)}  &  \underline{1.0292}  & 0.6864  &  1.2870 & 0.8531 & \underline{1.0245} &  \bf 0.6931   \\
\bottomrule
\end{tabular}
\label{tab:rating}
\vspace{-10pt}
\end{table}

\begin{table*}[ht!]
\centering
\caption{Performance comparison on sequential recommendation.}
\vspace{-10pt}
\begin{adjustbox}{width=0.98\linewidth}
\begin{tabular}{ccccccccccccc}
\toprule
\multirow{2.5}{*}{Methods} & \multicolumn{4}{c}{\textbf{Sports}} & \multicolumn{4}{c}{\textbf{Beauty}} &  \multicolumn{4}{c}{\textbf{Toys}} \\
\cmidrule(lr){2-5}\cmidrule(lr){6-9}\cmidrule(lr){10-13}
 & HR@5  & NDCG@5 & HR@10  & NDCG@10 & HR@5  & NDCG@5 & HR@10  & NDCG@10 & HR@5  & NDCG@5 & HR@10  & NDCG@10  \\
\cmidrule{1-13}
Caser   & 0.0116  & 0.0072  & 0.0194 & 0.0097 & 0.0205 & 0.0131 & 0.0347 & 0.0176 & 0.0166 & 0.0107 & 0.0270 & 0.0141  \\
HGN    &  0.0189  & 0.0120  & 0.0313  &  0.0159 & 0.0325  & 0.0206  & 0.0512  & 0.0266  & 0.0321  & 0.0221  & 0.0497  & 0.0277  \\
GRU4Rec   & 0.0129  & 0.0086  & 0.0204  & 0.0110  & 0.0164  & 0.0099  & 0.0283  & 0.0137  & 0.0097  & 0.0059  & 0.0176  & 0.0084 \\
BERT4Rec   & 0.0115   & 0.0075  & 0.0191  &  0.0099 &  0.0203 & 0.0124  & 0.0347  & 0.0170  & 0.0116   & 0.0071  & 0.0203  & 0.0099 \\
FDSA    &  0.0182  & 0.0122  & 0.0288  & 0.0156  & 0.0267  & 0.0163  & 0.0407  & 0.0208  & 0.0228  & 0.0140  & 0.0381  & 0.0189 \\
SASRec &  0.0233  &  0.0154 & 0.0350  &  0.0192 & 0.0387  & 0.0249  & 0.0605  & 0.0318  &  0.0463  &  0.0306  &  0.0675  & 0.0374 \\
S$^3$-Rec & 0.0251  & 0.0161  &  0.0385 & 0.0204 & 0.0387 & 0.0244  & 0.0647  & 0.0327  & 0.0443  & 0.0294  &  0.0700  &   0.0376 \\
P5-S {\color{purple}(2-3)}   & 0.0272   & 0.0169  & 0.0361  &  0.0198  &  \underline{0.0503}  & \underline{0.0370}  & \underline{0.0659}  & \underline{0.0421}  &  \bf 0.0648  & \bf 0.0567  & \bf 0.0709   & \bf 0.0587   \\
P5-B {\color{purple}(2-3)} &  \underline{0.0364}   & \underline{0.0296}   & \underline{0.0431}  & \underline{0.0318}  & \bf 0.0508  & \bf 0.0379  & \bf 0.0664 & \bf 0.0429  & 0.0608  & 0.0507  & 0.0688  &  0.0534  \\
P5-S {\color{orange}(2-13)}  & 0.0258   & 0.0159  & 0.0346  &  0.0188  &  0.0490   & 0.0358  & 0.0646  & 0.0409  &  \underline{0.0647}  & \underline{0.0566} &  \underline{0.0705} &  \underline{0.0585}  \\
P5-B {\color{orange}(2-13)}  & \bf 0.0387   & \bf 0.0312  & \bf 0.0460  &  \bf 0.0336  &  0.0493  & 0.0367  & 0.0645  & 0.0416  &  0.0587  &  0.0486  & 0.0675  & 0.0536    \\
\bottomrule
\end{tabular}
\end{adjustbox}
\vspace{-5pt}
\label{tab:sequential}
\end{table*}

\begin{table*}[ht!]
\centering
\caption{Performance comparison on explanation generation (\%).}
\vspace{-10pt}
\begin{adjustbox}{width=0.97\linewidth}
\begin{tabular}{ccccccccccccc}
\toprule
\multirow{2.5}{*}{Methods} & \multicolumn{4}{c}{\textbf{Sports}} & \multicolumn{4}{c}{\textbf{Beauty}} &  \multicolumn{4}{c}{\textbf{Toys}} \\
\cmidrule(lr){2-5}\cmidrule(lr){6-9}\cmidrule(lr){10-13}
 & BLUE4  & ROUGE1 & ROUGE2  & ROUGEL & BLUE4  & ROUGE1 & ROUGE2  & ROUGEL & BLUE4  & ROUGE1 & ROUGE2  & ROUGEL  \\
\cmidrule{1-13}
Attn2Seq   & 0.5305  & 12.2800  & 1.2107 & 9.1312 & 0.7889 & 12.6590 & 1.6820 & 9.7481 & 1.6238 & 13.2245 & 2.9942 & 10.7398  \\
NRT   &  0.4793  & 11.0723  & 1.1304  & 7.6674 & 0.8295  & 12.7815  & 1.8543  & 9.9477  & 1.9084  & 13.5231  & 3.6708  & 11.1867  \\
PETER   & 0.7112  & 12.8944  & 1.3283  & 9.8635  & \underline{1.1541}  & 14.8497  & \underline{2.1413}  & 11.4143  & 1.9861  & 14.2716  & 3.6718  & 11.7010 \\
P5-S {\color{purple}(3-3)}  &  \bf 1.0447   & \bf 14.9048  & \bf 2.1297  &  \bf 11.1778  & \bf 1.2237  & \bf 17.6938  & \bf 2.2489  &  \bf 12.8606 &  \underline{2.2892} & \bf 15.4505  & \underline{3.6974}  & \bf 12.1718  \\
P5-B {\color{purple}(3-3)} &  \underline{1.0407}   &  \underline{14.1589}   & \underline{2.1220}  & \underline{10.6096}  & 0.9742  & \underline{16.4530}  & 1.8858  & \underline{11.8765} & \bf 2.3185  & \underline{15.3474}  & \bf 3.7209 & \underline{12.1312} \\
\cmidrule{1-13}
PETER+  & \bf 2.4627 & \bf 24.1181  & 5.1937  & \bf 18.4105  & \bf 3.2606  & \underline{25.5541}  &  5.9668  &  \underline{19.7168}  & \bf 4.7919  & \underline{28.3083}  & 9.4520  & \bf 22.7017 \\
P5-S {\color{purple}(3-9)} & 1.4101   & \underline{23.5619}  & \bf 5.4196  &  \underline{17.6245}  & \underline{1.9788}  &  \bf 25.6253  &  \bf 6.3678  & \bf 19.9497  & 4.1222  &  \bf 28.4088  & 9.5432 &  \underline{22.6064} \\
P5-B {\color{purple}(3-9)} &  \underline{1.4689}  &  23.5476  & \underline{5.3926}  & 17.5852   & 1.8765  &  25.1183 & 6.0764 & 19.4488  &  3.8933 &  27.9916  & \underline{9.5896} &  22.2178 \\
P5-S {\color{orange}(3-12)}  & 1.3212   & 23.2474  & 5.3461   &  17.3780 & 1.9425 &  25.1474  & 6.0551  &  19.5601 & \underline{4.2764}  & 28.1897  & 9.1327 & 22.2514  \\
P5-B {\color{orange}(3-12)} &  1.4303  & 23.3810  &  5.3239  & 17.4913   &  1.9031 &  25.1763 & \underline{6.1980}  & 19.5188  & 3.5861  & 28.1369  & \bf 9.7562 & 22.3056 \\
\bottomrule
\end{tabular}
\end{adjustbox}
\vspace{-5pt}
\label{tab:explanation}
\end{table*}

\subsection{Performance Comparison on Different Task Families (RQ1)}
\label{sec:rq1}
In this section, we pretrain P5 with prompts from all five task families to verify its multitask learning ability. According to the default pretrain–predict task combination, we leave Prompt 1-10, Prompt 2-13, Prompt 3-12, Prompt 4-4, and Prompt 5-8 for zero-shot evaluation and pretrain P5 with the remaining personalized prompts. The performances of P5 and relevant baselines on the five task families are presented in Table~\ref{tab:rating} to Table~\ref{tab:direct}. For each task family, we choose one or more {\color{purple} \bf seen} prompts as supplement to the aforementioned zero-shot {\color{orange} \bf unseen} prompts to perform evaluations.

\subsubsection{\textbf{Rating Prediction}} Prompt 1-6 and Prompt 1-10 are used for evaluating P5's performance on rating prediction. The performance comparison is presented in Table~\ref{tab:rating}. We can see that when testing with seen Prompt 1-6, P5-B gets better MAE and slightly higher RMSE on all three datasets compared with MF. When testing with unseen Prompt 1-10, P5-B can achieve similar performance as Prompt 1-6. Moreover, P5-S usually has better MAE but higher RMSE. It seems that P5 is overfitting these data since the task complexity of rating prediction is relatively lower than other recommendation tasks. Overall, these results show that it is feasible to perform rating prediction on a conditional text generation framework.

\subsubsection{\textbf{Sequential Recommendation}} As illustrated in Table~\ref{tab:sequential}, Prompt 2-3 and Prompt 2-13 are employed for the evaluation of sequential recommendation under all-item setting, i.e., using all items as candidates rather than sampling 100 or 1,000 items for ranking. From the table, we can see that P5-B surpasses all competitive baselines with a relatively large gap on both seen (Prompt 2-3) and unseen (Prompt 2-13) prompts. On \emph{Toys}, P5-S can get even better performance than P5-B. While on \emph{Beauty} and \emph{Sports}, P5-B achieves the advantage over P5-S. The results show that the P5 architecture is effective in modeling the user interaction history and conducting next item prediction with the help of beam search.

\begin{table}[t!]
\centering
\footnotesize
\caption{Performance on review preference prediction.}
\vspace{-10pt}
\begin{adjustbox}{width=0.95\linewidth}
\begin{tabular}{ccccccc}
\toprule
\multirow{2.5}{*}{Methods} & \multicolumn{2}{c}{\textbf{Sports}} & \multicolumn{2}{c}{\textbf{Beauty}} &  \multicolumn{2}{c}{\textbf{Toys}} \\
\cmidrule(lr){2-3}\cmidrule(lr){4-5}\cmidrule(lr){6-7}
 & RMSE  & MAE & RMSE  & MAE & RMSE  & MAE \\
\cmidrule{1-7}
T0 {\color{purple}(4-2)}  &  0.6728  & 0.3140 & 0.6925  & 0.3324  & 0.8282  & 0.4201    \\
T0 {\color{orange}(4-4)}  &  \underline{0.6503}  & 0.2984  &   0.7066 &  0.3663 & 0.8148 & 0.4230    \\
P5-S {\color{purple}(4-2)}  &  0.7293  & 0.3529  & \bf 0.6233  & \bf 0.3051  & \bf 0.6464  & \bf 0.3125  \\
P5-B {\color{purple}(4-2)}   &  \bf 0.6487  & \bf 0.2847  & 0.6449  & 0.3168  & 0.6785  &  0.3342   \\
P5-S {\color{orange}(4-4)}   &  0.7565  & 0.3395  & \underline{0.6262}  & 0.3113  &  \underline{0.6577}  &  \underline{0.3174}  \\
P5-B {\color{orange}(4-4)}   & 0.6563  & \underline{0.2921}  & 0.6515 & \underline{0.3106} & 0.6730 & 0.3342  \\
\bottomrule
\end{tabular}
\end{adjustbox}
\label{tab:emotional}
\vspace{-10pt}
\end{table}

\subsubsection{\textbf{Explanation Generation}} In Table~\ref{tab:explanation}, Prompt 3-9 and Prompt 3-12 are used to evaluate P5's performance on explanation generation under feature-based setup, while Prompt 3-3 is used for direct explanation generation without providing a hint word. 
We can see that for Prompt 3-3, P5 achieves the best performances against all baselines. 
For feature-based prompts (Prompts 3-9 \& 3-12), P5 can outperform PETER+ on most cases, especially for \emph{Beauty} and \emph{Toys}.

\begin{table*}[ht!]
\centering
\caption{Performance comparison on review summarization (\%).}
\vspace{-10pt}
\begin{adjustbox}{width=1.0\linewidth}
\begin{tabular}{ccccccccccccc}
\toprule
\multirow{2.5}{*}{Methods} & \multicolumn{4}{c}{\textbf{Sports}} & \multicolumn{4}{c}{\textbf{Beauty}} &  \multicolumn{4}{c}{\textbf{Toys}} \\
\cmidrule(lr){2-5}\cmidrule(lr){6-9}\cmidrule(lr){10-13}
 & BLUE2  & ROUGE1 & ROUGE2  & ROUGEL & BLUE2  & ROUGE1 & ROUGE2  & ROUGEL & BLUE2  & ROUGE1 & ROUGE2  & ROUGEL  \\
\cmidrule{1-13}
T0 {\color{purple}(4-1)}  & 2.1581 & 2.2695  & 0.5694 & 1.6221 & 1.2871 & 1.2750 & 0.3904 & 0.9592 & \underline{2.2296} & 2.4671 & 0.6482 & 1.8424  \\
GPT-2 {\color{purple}(4-1)} &  0.7779  &  4.4534  & 1.0033 &  1.9236 &  0.5879 &  3.3844 &  0.6756 &  1.3956 &  0.6221 &  3.7149 &  0.6629 &  1.4813  \\
P5-S {\color{purple}(4-1)}  &  \underline{2.4962}  & \underline{11.6701}  & \underline{2.7187} & \underline{10.4819} & \bf 2.1225 & \bf 8.4205 & \bf 1.6676 & \bf 7.5476 & \bf 2.4752 & \bf 9.4200 & \bf 1.5975 & \bf 8.2618  \\
P5-B {\color{purple}(4-1)} & \bf 2.6910 & \bf 12.0314  & \bf 3.2921 &  \bf 10.7274 & \underline{1.9325} & \underline{8.2909} & \underline{1.4321}  & \underline{7.4000} & 1.7833 &  \underline{8.7222} & \underline{1.3210} & \underline{7.6134}  \\
\bottomrule
\end{tabular}
\end{adjustbox}
\label{tab:summarize}
\end{table*}

\begin{table*}[ht!]
\centering
\caption{Performance comparison on direct recommendation.}
\vspace{-10pt}
\begin{adjustbox}{width=1.0\linewidth}
\begin{tabular}{cccccccccccccccc}
\toprule
\multirow{2.5}{*}{Methods} & \multicolumn{5}{c}{\textbf{Sports}} & \multicolumn{5}{c}{\textbf{Beauty}} &  \multicolumn{5}{c}{\textbf{Toys}} \\
\cmidrule(lr){2-6}\cmidrule(lr){7-11}\cmidrule(lr){12-16}
 & HR@1 & HR@5  & NDCG@5 & HR@10  & NDCG@10 & HR@1 &  HR@5  & NDCG@5 & HR@10  & NDCG@10 & HR@1 &  HR@5  & NDCG@5 & HR@10  & NDCG@10  \\
\cmidrule{1-16}
BPR-MF  & 0.0314   & 0.1404  & 0.0848 &  0.2563 &  0.1220 & 0.0311  & 0.1426  & 0.0857  & 0.2573  & 0.1224  &  0.0233  & 0.1066  & 0.0641  & 0.2003  &  0.0940 \\
BPR-MLP &  0.0351 & 0.1520  & 0.0927 & 0.2671 & 0.1296  & 0.0317 &  0.1392 &  0.0848 &  0.2542 & 0.1215  & 0.0252  & 0.1142  & 0.0688  & 0.2077  & 0.0988  \\
SimpleX  & 0.0331   &  \bf 0.2362   & \bf 0.1505  &  \underline{0.3290}   &  \underline{0.1800}   &  0.0325  &  \underline{0.2247}   &  \underline{0.1441}  &  0.3090  &  \underline{0.1711}  & 0.0268   & \bf 0.1958   &  \bf 0.1244   &  \bf 0.2662   & \bf 0.1469  \\
\cmidrule{1-16}
P5-S {\color{purple}(5-1)}  &  0.0638  &  0.2096   &  0.1375  &  0.3143    &   0.1711   &  0.0600  &  0.2021   &  0.1316  &  \underline{0.3121}  &  0.1670  &  0.0405  &  \underline{0.1538}  &  \underline{0.0969}  &  \underline{0.2405}  &  \underline{0.1248}  \\
P5-B {\color{purple}(5-1)}  &  0.0245  &   0.0816  &  0.0529  & 0.1384  &  0.0711   &  0.0224  &  0.0904   &  0.0559  &  0.1593  &  0.0780  & 0.0187  &  0.0827  & 0.0500   &  0.1543  &  0.0729 \\
P5-S {\color{purple}(5-4)}  &  \underline{0.0701}  &   \underline{0.2241}   &  \underline{0.1483}  &  \bf 0.3313   & \bf 0.1827   &  \bf 0.0862  &  \bf 0.2448   &  \bf 0.1673   & \bf 0.3441   & \bf 0.1993  &  0.0413  &  0.1411   &  0.0916  &  0.2227   &  0.1178 \\
P5-B {\color{purple}(5-4)}  &  0.0299   &  0.1026   &  0.0665   &  0.1708   &  0.0883   &  0.0506  &  0.1557   &  0.1033  &  0.2350  &  0.1287  &  0.0435  &  0.1316   &  0.0882  &   0.2000  &  0.1102 \\
\cmidrule{1-16}
P5-S {\color{purple}(5-5)}  & 0.0574   &  0.1503  & 0.1050  &  0.2207  & 0.1276  & 0.0601 & 0.1611  & 0.1117  & 0.2370  & 0.1360 &  \underline{0.0440}  & 0.1282   & 0.0865 &  0.2011 &  0.1098  \\
P5-B {\color{purple}(5-5)}  & 0.0641   &  0.1794  & 0.1229  &  0.2598  & 0.1488  & 0.0588 & 0.1573 & 0.1089 & 0.2325 & 0.1330  & 0.0386   & 0.1122 &  0.0756  &  0.1807 & 0.0975  \\
P5-S {\color{orange} (5-8)} & 0.0567   &  0.1514  & 0.1049  &  0.2196  & 0.1269  & 0.0571 & 0.1566 & 0.1078 & 0.2317 & 0.1318  & \bf 0.0451  & 0.1322  & 0.0889 &  0.2023 & 0.1114  \\
P5-B {\color{orange} (5-8)} & \bf 0.0726  &  0.1955  & 0.1355  &  0.2802  & 0.1627  & \underline{0.0608} & 0.1564 & 0.1096 & 0.2300 & 0.1332  &  0.0389  & 0.1147  &  0.0767 &  0.1863  &  0.0997  \\
\bottomrule
\end{tabular}
\end{adjustbox}
\vspace{-5pt}
\label{tab:direct}
\end{table*}

\subsubsection{\textbf{Review Related}} We take Prompts 4-2 and 4-4 to compare P5's performance with T0 on review preference prediction, as shown in Table~\ref{tab:emotional}. We can see that P5-S achieves better RMSE and MAE on \emph{Beauty} and \emph{Toys}, while P5-B shows better performance on \emph{Sports}. Additionally, we take Prompt 4-1 to evaluate P5's ability on review summarization, as shown in Table~\ref{tab:summarize}. 
For this task, P5-S clearly outperforms T0 and GPT-2 on both \emph{Beauty} and \emph{Toys} datasets.
It is worth noting that GPT-2 and T0 has 1.5B and 11B parameters, respectively. 
This shows that P5 can achieve better 
performances than these competitive baselines with a much smaller model size.

\subsubsection{\textbf{Direct Recommendation}} Finally, Prompts 5-1, 5-4, 5-5 and 5-8 are applied to evaluate the direct recommendation task under the 1-out-of-100 evaluation setting.
For binary question prompts (5-1 \& 5-4), which are discriminative prompts, we use the softmax generation probability of ``yes'' to rank the candidate items. For open question prompts (5-5 \& 5-8), which are generative prompts, we use beam-search (Eq.\eqref{eq:2}) to generate the top-$k$ list.
The results are presented in Table~\ref{tab:direct}. From the table, we can see that P5-B and P5-S have great advantages over BPR-MF and BPR-MLP on all three datasets. 
Comparing with SimpleX, we can see that P5 works especially well on top-1 item ranking, which is more than two times better than SimpleX on HR@1. 
Besides, P5 also achieves the best result on most of the other metrics.
The success of P5 on direct recommendation shows the competence of the sequence-to-sequence generation framework in recommendation domain.

\subsection{Zero-shot Generalization to Unseen Prompts and Items in New Domain (RQ2)}

\subsubsection{\textbf{Transfer to Unseen Personalized Prompts}}
In this section, we transfer the pretrained P5 models to the previously held-out prompts during pretraining. These unseen prompts are from the same task families, and the testing items have been seen by P5 during pretraining at least once. The experimental results are also reported in Table~\ref{tab:rating} to Table~\ref{tab:direct}. As previously discussed in Section~\ref{sec:rq1}, P5 achieves surprisingly good performances on various task families when being challenged by unseen prompts. 
On some specific datasets, the performances of P5 on unseen prompts even surpass seen prompts, e.g., P5-B gets the best performance under Prompt 2-13 on~\emph{Sports}. These results show that multitask prompted pretraining empowers P5 enough robustness to understand unseen prompts with wording variations.

\begin{table}[t!]
\centering
\caption{Statistics on domain transfer evaluation sets.}
\vspace{-10pt}
\begin{tabular}{lrrrr}
\toprule
Dataset &  {\textbf{Sports}} &  {\textbf{Beauty}} & {\textbf{Toys}}\\
\cmidrule{1-4}
\#Users    &  290   &  439   &  487    \\
\#Items    &  381   &  586  & 886  \\
\#Reviews  & 478   &  1,237   & 1,183  \\
\bottomrule
\end{tabular}
\label{tab:zero-stats}
\vspace{-10pt}
\end{table}

\begin{figure*}[t!]
 \centering
 \includegraphics[width=\linewidth]{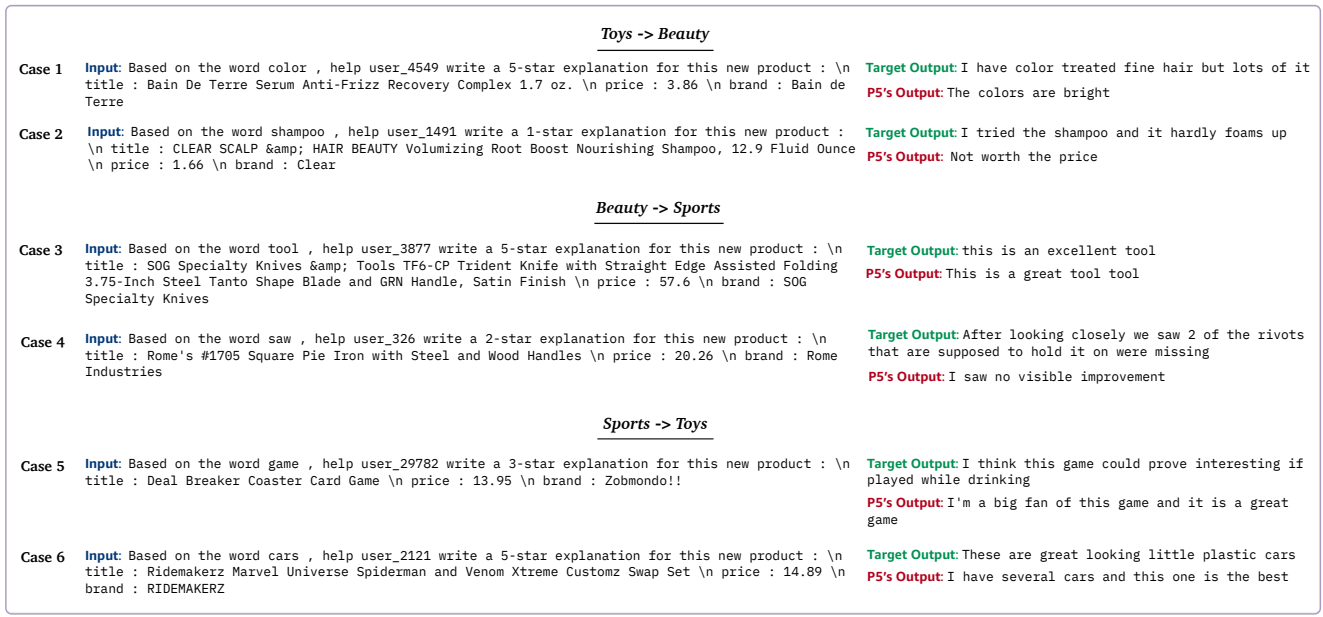}
 \vspace{-20pt}
 \caption{Example cases of zero-shot domain transfer on {\color{orange} Z-6} task. We demonstrate three transfer directions: \emph{Toys} to \emph{Beauty}, \emph{Beauty} to \emph{Sports}, and \emph{Sports} to \emph{Toys}.}
\label{fig:case}
\vspace{-10pt}
\end{figure*}

\begin{figure*}[t!]
\centering
\includegraphics[width=\linewidth]{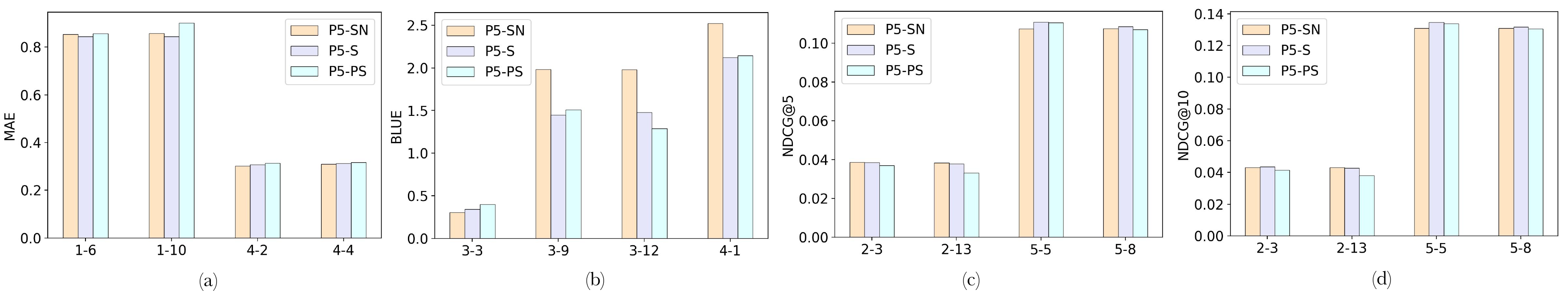}
\vspace{-20pt}
\caption{Performance comparison among P5-S, P5-SN, and P5-PS on \emph{Beauty}.}
\label{fig:ablation}
\vspace{-5pt}
\end{figure*}

\subsubsection{\textbf{Transfer to Items in New Domain}}
Next, we increase the difficulty level of zero-shot transfer. We collect a group of 741 users that exist in all the three domains with their interaction and review histories in other domains. The detailed statistics of these domain transfer evaluation sets are illustrated in Table~\ref{tab:zero-stats}. We then challenge P5-B pretrained on one domain with unseen prompts from the Task Family Z, whose item fields are filled with the information from a new product domain. For example, we ask the P5 model pretrained on the \emph{Toys} domain about an existing user's preference towards an item in the \emph{Beauty} domain. The full results on all six directions are reported in Table~\ref{tab:transfer}. From the table, we notice P5 still maintains sufficient performances for rating prediction (Prompts Z-2 \& Z-3), like/dislike prediction (Prompts Z-1 \& Z-4), as well as explanation generation with feature word (Prompt Z-6). In contrast, direct explanation generation without feature word (Prompts Z-5 \& Z-7) is very difficult for P5 because it lacks awareness of relevant knowledge in the new domain. In Figure~\ref{fig:case}, we provide some example explanations generated by P5-B under the setup of zero-shot domain transfer (Prompt Z-6). We can see that P5 is able to catch different users' rating preferences and hint feature words, then integrate them with the knowledge learned from previous domain to generate plausible explanations. 

\subsection{Ablation on Model Size (RQ3)}
In this section, we will discuss the influence of model size on the performance of P5 on different recommendation tasks. Here, we train two size variants of P5, namely P5-small and P5-base. The parameter numbers of these two P5 models are 60.75M and 223.28M, respectively. From Table~\ref{tab:rating} to Table~\ref{tab:direct}, we can see that although P5-S is only 1/4 of the size of P5-B, P5-S can beats P5-B on a series of tasks and datasets. For example, P5-S achieves better sequential recommendation, review preference prediction, and direct recommendation (Prompts 5-5 \& 5-8) performances than P5-B on \emph{Toys}. 
In contrast, P5-B shows advantages on sequential recommendation and review preference prediction tasks for \emph{Sports}.
Since \emph{Sports} contains more users, items and reviews and has a lower sparsity, it requires a model with higher capacity to discover latent correlation among different personalized factors. The findings indicate that larger P5 models may be needed when the dataset is large, while for smaller datasets, smaller P5 models could be enough. As a result, we should decide an appropriate model size that matches the scale of the training data.

\subsection{Ablation on Task Scaling (RQ3)}
Moreover, we explore whether multitask prompted pretraining is superior than pretraining on each task family alone. We pretrain P5-small on \emph{Beauty} dataset with prompts from every single task family, resulting in five models -- P5-S1, P5-S2, P5-S3,  P5-S4, and P5-S5. We then compare P5-S on various recommendation tasks with the corresponding single task P5 model. The performance comparison between P5-S and \textbf{P5-SN} ($N \in \left[1,2,3,4,5 \right]$) is illustrated in Figure~\ref{fig:ablation}. 
As shown in the figure, P5-S achieves comparable or better performance than P5-SN on rating prediction, sequential recommendation and direct recommendation tasks, while on text generation tasks such as explanation generation (Prompts 3-9 \& 3-12) and review summarization (Prompt 4-1), P5-SN is better than P5-S.
This indicates that multitask modeling (P5-S) seeks a good balance among tasks and improves recommendation performance by leveraging the power of language understanding.
Besides, both P5-S and P5-SN perform better than or comparable with state-of-the-art baselines on all tasks, as shown in Table \ref{tab:rating} through Table \ref{tab:direct}, which demonstrates the power of P5 for recommendation.

\begin{table}[t!]
\centering
\small
\caption{Performance on zero-shot domain transfer.}
\vspace{-10pt}
\setlength{\tabcolsep}{1pt}
\begin{tabular}{cccccccc}
\toprule
\multirow{2.5}{*}{Directions} & \multicolumn{1}{c}{\textbf{\color{orange}\footnotesize Z-1 \& Z-4}} & \multicolumn{1}{c}{\textbf{\color{orange}\footnotesize Z-2 \& Z-3}} &  \multicolumn{2}{c}{\textbf{\color{orange}\footnotesize Z-5 \& Z-7 (\%)}} & \multicolumn{2}{c}{\textbf{\color{orange}\footnotesize Z-6 (\%)}}\\
\cmidrule(lr){2-2}\cmidrule(lr){3-3}\cmidrule(lr){4-5}\cmidrule(lr){6-7}
 & Accuracy  & MAE &  BLUE2 & ROUGE1 & BLUE2  & ROUGE1 \\
\cmidrule{1-7}
\emph{\textbf{Toys} -> \textbf{Beauty}}   &  0.7922  & 0.8244  & 0.1940   & 3.5441  & 0.8623  & 13.8487 \\
\emph{\textbf{Toys} -> \textbf{Sports}}   &  0.8682  & 0.6644  & 0.1203   & 3.7684  & 0.3972  & 13.7654  \\
\cmidrule{1-7}
\emph{\textbf{Beauty} -> \textbf{Toys}}   &  0.8073  & 0.7792  & 0.0309   & 1.4904  & 2.7606  & 15.1632  \\
\emph{\textbf{Beauty} -> \textbf{Sports}}   &  0.8676  & 0.6838  & 0.0264   & 1.7033  & 1.6530  & 13.9460  \\
\cmidrule{1-7}
\emph{\textbf{Sports} -> \textbf{Toys}}   &  0.8230  & 0.7443  & 0.0060   & 1.7313  & 4.2334  & 17.1248  \\
\emph{\textbf{Sports} -> \textbf{Beauty}}  &  0.8057  & 0.8102 & 0.0080 & 2.0195 & 3.5059 & 18.5577 \\
\bottomrule
\end{tabular}
\label{tab:transfer}
\vspace{-10pt}
\end{table}

\begin{figure*}[t!]
\centering
\includegraphics[width=\linewidth]{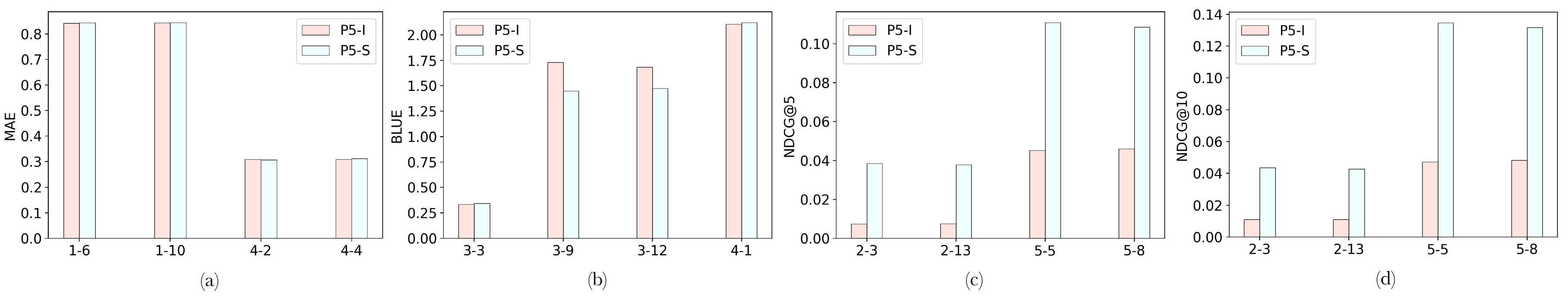}
\vspace{-20pt}
\caption{Performance of P5-S and P5-I on \emph{Beauty} showing the influence of how to implement personalization.}
\label{fig:personalization}
\end{figure*}

\subsection{Ablation on Prompt Scaling (RQ3)}
As mentioned in implementation details, our default pretrain--predict task combination follows the leave-one-out strategy. However, do we need so many prompts during pretraining to enable P5's zero-shot generalization ability? In this section, we explore to reduce the number of pretraining prompts and then make comparisons with the P5 model pretrained under default setup. To this end, we choose a collection of pretraining prompts that has the minimum number of prompts to cover all important personalized fields. Specifically, this combination contains the following 18 personalized prompts: \{1-5, 1-6, 1-8, 1-9, 2-1, 2-3, 2-8, 2-11, 3-2, 3-3, 3-6, 3-9, 4-1, 4-2, 4-3, 5-2, 5-5, 5-7\}. Similar to the default pretrain–predict combination, the last prompt in each task family is for zero-shot evaluation. We name this prompt scaling variant of P5-small as \textbf{P5-PS} and then pretrain P5-PS on \emph{Beauty} dataset. The performance comparison between P5-S and P5-PS is also presented in Figure~\ref{fig:ablation}. From the figure, we can observe that P5-S beats P5-PS on most tasks except for some generation tasks (i.e., Prompts 3-3, 3-9 \& 4-1). Interestingly, P5-S outperforms P5-PS on Prompt 3-12 -- a zero-shot explanation generation task. In fact, P5-S also shows its superiority on other zero-shot tasks such as Prompts 1-10, 2-13, and 5-8. Overall, we can find that larger number of high quality personalized prompts can generally help P5 achieve better performances on various recommendation tasks especially zero-shot tasks with unseen prompts.

\vspace{-3pt}
\subsection{How to Implement Personalization (RQ4)}
\label{sec:rq4}
In this section, we discuss different strategies to implement personalization in P5. The default practice is using SentencePiece tokenizer to split personalized fields into multiple sub-word units and meanwhile using whole-word embedding to preserve the field information (Figure \ref{fig:framework}). A straightforward alternative is creating an independent extra token for each user and item. Here we name this P5-small variant as \textbf{P5-I} and also pretrain it on \emph{Beauty} dataset. While the former utilizes collaborative learning to implicitly optimize the latent correlations among different sub-word tokens, the latter learns a unique personalized representation for every extra token. The performance comparison between P5-S and P5-I is shown in Figure~\ref{fig:personalization}. We can see that P5-I achieves similar performances as P5-S on regression tasks (Prompts 1-6 \& 1-10 for rating prediction, Prompts 4-2 \& 4-4 for review-based rating regression) and review summarization tasks (Prompt 4-1). Also, P5-I is slightly better than P5-S on explanation generation tasks (Prompts 3-3, 3-9 \& 3-12).
However, P5-I significantly underperforms P5-S by a large margin on both sequential and direct recommendation tasks (all prompts in Figure \ref{fig:personalization} (c) \& (d)). The reason behind P5-I's lower performance lies in that the newly introduced huge number of extra tokens and embeddings cannot be well trained compared with the original sub-word units initialized from T5. This shows that our default setting can achieve better recommendation and overall performances with the help of  collaborative learning while keeping a small and constant amount of learnable tokens.

\vspace{-2ex}
\section{Conclusions and Future Work}
In this paper, we present P5 which unifies different recommendation tasks into a shared language modeling and natural language generation framework. By designing a collection of personalized prompts covering five recommendation task families, we transfer all raw data such as the user-item interactions, user descriptions, item metadata, and user reviews to the same format -- input-target text pairs. We then pretrain P5 in a full language environment to help it discover deeper semantics for various recommendation tasks. According to our experiments, P5 can beat or achieve similar performance with several representative approaches on all five task families. Moreover, P5 shows the generalization ability on performing zero-shot transfer to new items, new domains, and new personalized prompts. 
In the future, we will continue exploring to further enlarge the model size of P5 and employ more powerful base models such as GPT-3, OPT, and BLOOM. Besides, P5 is a very flexible paradigm and it is promising to further extend P5 to diverse modalities and more tasks such as conversational recommendation, comparative recommendation, cross-platform recommendation, or even various search tasks by incorporating user queries into P5.
Finally, in this work, we designed explicit prompts since they are intuitive, flexible, and close to the natural way of how humans communicate with each other, which enables instruction-based recommendation, while in the future, we will also investigate prompt search and/or latent prompt techniques to achieve instruction prompts or leverage retrieval-enhanced generation to further boost P5's performance on downstream tasks.

\section*{Acknowledgment}
This work was supported in part by NSF IIS 1910154, 2007907, and 2046457. Any opinions, findings, conclusions or recommendations expressed in this material are those of the authors and do not necessarily reflect those of the sponsors.

\bibliographystyle{ACM-Reference-Format}
\bibliography{p5}

\clearpage
\newpage



\section*{Supplementary material}

# APPENDIX

In this appendix, we first provide additional experimental results on *Yelp* dataset in Section A. Then we collect and report statistics on the training and inference time of P5 variants in Section B. Since personalized prompts constitute a very important part of our work, we thus offer the full list of the personalized prompts for Amazon datasets in Section D and Yelp dataset in Section E, respectively.

## A  EXPERIMENTAL RESULTS ON *YELP* DATASET

For *Yelp* dataset, we also follow the default pretrain–predict task combination setup. Based on the personalized prompts presented in Section E, we leave Prompt 1-10, Prompt 2-13, Prompt 3-10, Prompt 4-3, and Prompt 5-8 for zero-shot evaluation and pretrain a P5-S model with the remaining prompts. We adopt the same baselines as described in Section 5.2 for performance comparison. Again, for each task family, we choose one or more **seen** prompts as supplement to the aforementioned zero-shot **unseen** prompts to perform evaluations. The performances of P5-S and relevant baseline models are shown in Table 10 to Table 14. As indicated in these experimental results, P5 also shows great capability on Yelp, especially for the sequential recommendation and explanation generation tasks.

## B  STATISTICS ON TRAINING & INFERENCE TIME (RQ5)

In this section, we provide statistics on the training and inference time of P5 models, we collect the running time on the *Beauty* dataset. As mentioned in Section 5.1, we trained our P5 models on 4 × A5000 GPUs. From our records, P5-S spent 6.7 hours to finish training, while P5-B took 24.3 hours due to the larger number of parameters. In terms of inference, we use a single A5000 GPU to conduct evaluations. The average inference time of P5-S and P5-B on different tasks are presented in Table 15. Among all tasks, sequential and direct recommendation tasks require much longer inference time than other tasks. This can be ascribed to the beam search step described in Eq. (2). Since we need to generate a list of recommended items, the larger the beam size $B$ is, the longer the decoding will take. Besides, we can also observe that direct recommendation typically takes longer than sequential recommendation. The reason is that 100 candidates are included in the input of direct recommendation prompts, which usually have more tokens than that of sequential recommendation prompts. Overall, even though the pretraining of P5 takes hours to finish, but the inference is very fast. It is also promising to further reduce the training and inference time with the help of efficient Transformer techniques.

Table 10: Performance comparison on rating prediction.

| Methods | Yelp | |
|---|---|---|
| | RMSE | MAE |
| MF | **1.2645** | 1.0426 |
| MLP | 1.2951 | 1.0340 |
| P5-S (1-6) | 1.4868 | 1.0186 |
| P5-S (1-10) | 1.4685 | **1.0054** |

Table 11: Performance on sequential recommendation.

| Methods | Yelp | | | |
|---|---|---|---|---|
| | HR@5 | NDCG@5 | HR@10 | NDCG@10 |
| Caser | 0.0151 | 0.0096 | 0.0253 | 0.0129 |
| HGN | 0.0186 | 0.0115 | 0.0326 | 0.0159 |
| GRU4Rec | 0.0152 | 0.0099 | 0.0263 | 0.0134 |
| BERT4Rec | 0.0051 | 0.0033 | 0.0090 | 0.0045 |
| FDSA | 0.0158 | 0.0098 | 0.0276 | 0.0136 |
| SASRec | 0.0162 | 0.0100 | 0.0274 | 0.0136 |
| $S^3$-Rec | 0.0201 | 0.0123 | 0.0341 | 0.0168 |
| P5-S (2-3) | 0.0568 | 0.0402 | **0.0707** | **0.0447** |
| P5-S (2-13) | **0.0574** | **0.0403** | 0.0703 | 0.0445 |

Table 12: Performance on explanation generation (%).

| Methods | Yelp | | | |
|---|---|---|---|---|
| | BLUE4 | ROUGE1 | ROUGE2 | ROUGEL |
| Attn2Seq | 0.8031 | 14.1185 | 1.9730 | 10.9220 |
| NRT | 0.8128 | 13.9256 | 1.9635 | 10.6980 |
| PETER | 0.5938 | 12.0065 | 1.8645 | 9.4645 |
| P5-S (3-2) | **1.2717** | **18.1796** | **2.9477** | **13.5465** |
| PETER+ | **3.2827** | **27.2366** | **8.1941** | **21.1573** |
| P5-S (3-7) | 2.9797 | 27.1860 | 6.6827 | 19.6348 |
| P5-S (3-10) | 3.0600 | 27.2990 | 6.7655 | 19.6329 |

Table 13: Performance on review preference prediction.

| Methods | Yelp | |
|---|---|---|
| | RMSE | MAE |
| T0 (4-2) | 0.5383 | 0.2756 |
| T0 (4-3) | **0.5359** | **0.2732** |
| P5-S (4-2) | 0.6301 | 0.3113 |
| P5-S (4-3) | 0.6350 | 0.3150 |

Table 14: Performance on direct recommendation.

| Methods | Yelp | | | | |
|---|---|---|---|---|---|
| | HR@1 | HR@5 | NDCG@5 | HR@10 | NDCG@10 |
| BPR-MF | **0.0813** | 0.2251 | 0.1543 | 0.3312 | 0.1886 |
| BPR-MLP | 0.0489 | 0.1876 | 0.1184 | 0.3066 | 0.1566 |
| SimpleX | 0.0569 | **0.3970** | **0.2538** | **0.5473** | **0.3020** |
| P5-S (5-1) | 0.0603 | 0.2097 | 0.1354 | 0.3197 | 0.1708 |
| P5-S (5-4) | 0.0567 | 0.2065 | 0.1320 | 0.3190 | 0.1682 |
| P5-S (5-5) | 0.0696 | 0.2134 | 0.1423 | 0.3178 | 0.1758 |
| P5-S (5-8) | 0.0705 | 0.2136 | 0.1432 | 0.3219 | 0.1780 |

Table 15: Average inference time (in milliseconds) of P5 variants on different tasks on the *Beauty* dataset.

| | per user | | per user–item pair | | per review | |
|---|---|---|---|---|---|---|
| Models | Sequential | Direct | Rating | Explanation | Summarization | Preference |
| | (2-13) | (5-8) | (1-10) | (3-12) | (4-1) | (4-4) |
| P5-S | 53.96 | 57.40 | 4.93 | 13.41 | 8.66 | 6.25 |
| P5-B | 78.19 | 98.80 | 13.40 | 34.71 | 23.51 | 14.85 |

Table 16: Performance comparison on direct recommendation under larger sample size.

| Methods | Sports | | | | | Beauty | | | | | Toys | | | | |
|---|---|---|---|---|---|---|---|---|---|---|---|---|---|---|---|
| | HR@1 | HR@5 | NDCG@5 | HR@10 | NDCG@10 | HR@1 | HR@5 | NDCG@5 | HR@10 | NDCG@10 | HR@1 | HR@5 | NDCG@5 | HR@10 | NDCG@10 |
| BPR-MF | 0.0314 | 0.1404 | 0.0848 | 0.2563 | 0.1220 | 0.0311 | 0.1426 | 0.0857 | 0.2573 | 0.1224 | 0.0233 | 0.1066 | 0.0641 | 0.2003 | 0.0940 |
| BPR-MLP | 0.0351 | 0.1520 | 0.0927 | 0.2671 | 0.1296 | 0.0317 | 0.1392 | 0.0848 | 0.2542 | 0.1215 | 0.0252 | 0.1142 | 0.0688 | 0.2077 | 0.0988 |
| SimpleX | 0.0331 | 0.2362 | 0.1505 | 0.3290 | 0.1800 | 0.0325 | 0.2247 | 0.1441 | 0.3090 | 0.1711 | 0.0268 | **0.1958** | **0.1244** | 0.2662 | 0.1469 |
| P5-S (5-1) | 0.0638 | 0.2096 | 0.1375 | 0.3143 | 0.1711 | 0.0600 | 0.2021 | 0.1316 | 0.3121 | 0.1670 | 0.0405 | 0.1538 | 0.0969 | 0.2405 | 0.1248 |
| P5-B (5-1) | 0.0245 | 0.0816 | 0.0529 | 0.1384 | 0.0711 | 0.0224 | 0.0904 | 0.0559 | 0.1593 | 0.0780 | 0.0187 | 0.0827 | 0.0500 | 0.1543 | 0.0729 |
| P5-S (5-4) | 0.0701 | 0.2241 | 0.1483 | 0.3313 | 0.1827 | 0.0862 | 0.2448 | 0.1673 | 0.3441 | 0.1993 | 0.0413 | 0.1411 | 0.0916 | 0.2227 | 0.1178 |
| P5-B (5-4) | 0.0299 | 0.1026 | 0.0665 | 0.1708 | 0.0883 | 0.0506 | 0.1557 | 0.1033 | 0.2350 | 0.1287 | 0.0435 | 0.1316 | 0.0882 | 0.2000 | 0.1102 |
| P5-S (5-5) | 0.0766 | 0.2195 | 0.1495 | 0.3187 | 0.1814 | 0.0826 | 0.2378 | 0.1613 | 0.3289 | 0.1906 | 0.0453 | 0.1404 | 0.0933 | 0.2192 | 0.1185 |
| P5-B (5-5) | 0.0618 | 0.1812 | 0.1226 | 0.2669 | 0.1501 | 0.0601 | 0.1607 | 0.1113 | 0.2308 | 0.1339 | 0.0438 | 0.1367 | 0.0910 | 0.2106 | 0.1147 |
| P5-S (5-8) | **0.0967** | **0.2519** | **0.1763** | **0.3530** | **0.2089** | **0.1020** | **0.2688** | **0.1869** | **0.3665** | **0.2184** | **0.0588** | 0.1793 | 0.1203 | **0.2691** | **0.1491** |
| P5-B (5-8) | 0.0631 | 0.1834 | 0.1245 | 0.2643 | 0.1505 | 0.0576 | 0.1572 | 0.1086 | 0.2306 | 0.1321 | 0.0442 | 0.1374 | 0.0914 | 0.2160 | 0.1167 |

## C INFLUENCE OF SAMPLE SIZE

When creating the training sentences for direct recommendation under the generative prompts (Prompt 5-5 to 5-8), for each user-item interaction in the training set, we randomly sample 99 negative items that the user did not interact with. Together with the positive item, a 100-item candidate list is then created. We then randomly select a prompt from Prompt 5-5 to Prompt 5-8 to create a training sentence. For each user-item interaction, we repeat the above process for $N$ times and achieve $N$ training sentences. In the default setting, we set the sample size $N$ to 10. We find that increasing the training sample size could benefit the performance of direct recommendation under the generative prompts. We increase the sample size from the default number 10 to 200 and report the performances on Prompts 5-5 and 5-8 in Table 16. From the table, we can see that a larger data sample size can greatly boost the direct recommendation performances.

## D FULL LIST OF PERSONALIZED PROMPTS FOR *AMAZON* DATASETS

### D.1 Task Family 1: Rating Prediction

**Prompt ID: 1-1**

Input template: Which star rating will user_{{user_id}} give item_{{item_id}}? (1 being lowest and 5 being highest)

Target template: {{star_rating}}

**Prompt ID: 1-2**

Input template: How will user_{{user_id}} rate this product: {{item_title}}? (1 being lowest and 5 being highest)

Target template: {{star_rating}}

**Prompt ID: 1-3**

Input template: Will user_{{user_id}} give item_{{item_id}} a {{star_rating}}-star rating? (1 being lowest and 5 being highest)

Target template: {{answer_choices[label]}} (yes/no)

**Prompt ID: 1-4**

Input template: Does user_{{user_id}} like or dislike item_{{item_id}}?

Target template:
{{answer_choices[label]}} (like/dislike) – like (4,5) / dislike (1,2,3)

**Prompt ID: 1-5**

Input template: Predict the user_{{user_id}} 's preference on item_{{item_id}} ({{item_title}})
-1 \n -2 \n -3 \n -4 \n -5

Target template: {{star_rating}}

**Prompt ID: 1-6**

Input template: What star rating do you think {{user_desc}} will give item_{{item_id}}? (1 being lowest and 5 being highest)

Target template: {{star_rating}}

**Prompt ID: 1-7**

Input template: How will {{user_desc}} rate this product: {{item_title}}? (1 being lowest and 5 being highest)

Target template: {{star_rating}}

**Prompt ID: 1-8**

Input template: Will {{user_desc}} give a {{star_rating}}-star rating for {{item_title}}? (1 being lowest and 5 being highest)

Target template: {{answer_choices[label]}} (yes/no)

**Prompt ID: 1-9**

Input template: Does {{user_desc}} like or dislike {{item_title}}?

Target template:
{{answer_choices[label]}} (like/dislike) – like (4,5) / dislike (1,2,3)

**Prompt ID: 1-10**

Input template: Predict {{user_desc}} 's preference towards {{item_title}} (1 being lowest and 5 being highest)

Target template: {{star_rating}}

### D.2 Task Family 2: Sequential Recommendation

**Prompt ID: 2-1**

Input template: Given the following purchase history of user_{{user_id}}:
{{purchase_history}}
predict next possible item to be purchased by the user?

Target template: {{next_item}}

**Prompt ID: 2-2**

Input template: I find the purchase history list of user_{{user_id}}:
{{purchase_history}}
I wonder which is the next item to recommend to the user. Can you help me decide?

Target template: {{next_item}}

**Prompt ID: 2-3**

14

```
Input template: Here is the purchase history list of
user_{{user_id}}:
{{purchase_history}}
try to recommend next item to the user

Target template: {{next_item}}
```

### Prompt ID: 2-4

```
Input template: Given the following purchase history of
{{user_desc}}:
{{purchase_history}}
predict next possible item for the user

Target template: {{next_item}}
```

### Prompt ID: 2-5

```
Input template: Based on the purchase history of {{user_desc}}:
{{purchase_history}}
Can you decide the next item likely to be purchased by the user?

Target template: {{next_item}}
```

### Prompt ID: 2-6

```
Input template: Here is the purchase history of {{user_desc}}:
{{purchase_history}}
What to recommend next for the user?

Target template: {{next_item}}
```

### Prompt ID: 2-7

```
Input template: Here is the purchase history of user_{{user_id}}:
{{purchase_history}}
Select the next possible item likely to be purchased by the user from
the following candidates:
{{candidate_items}}

Target template: {{next_item}}
```

### Prompt ID: 2-8

```
Input template: Given the following purchase history of
{{user_desc}}:
{{purchase_history}}
What to recommend next for the user? Select one from the following
items:
{{candidate_items}}

Target template: {{next_item}}
```

### Prompt ID: 2-9

```
Input template: Based on the purchase history of user_{{user_id}}:
{{purchase_history}}
Choose the next possible purchased item from the following
candidates:
{{candidate_items}}

Target template: {{next_item}}
```

### Prompt ID: 2-10

```
Input template: I find the purchase history list of {{user_desc}}:
{{purchase_history}}
I wonder which is the next item to recommend to the user. Try to
select one from the following candidates:
{{candidate_items}}

Target template: {{next_item}}
```

### Prompt ID: 2-11

```
Input template: User_{{user_id}} has the following purchase history:
{{purchase_history}}
Does the user likely to buy {{candidate_item}} next?

Target template: {{answer_choices[label]}} (yes/no)
```

### Prompt ID: 2-12

```
Input template: According to {{user_desc}} 's purchase history list:
{{purchase_history}}
Predict whether the user will purchase {{candidate_item}} next?
```

```
Target template: {{answer_choices[label]}} (yes/no)
```

### Prompt ID: 2-13

```
Input template: According to the purchase history of {{user_desc}}:
{{purchase_history}}
Can you recommend the next possible item to the user?

Target template: {{next_item}}
```

## D.3   Task Family 3: Explanation Generation

### Prompt ID: 3-1

```
Input template: Generate an explanation for user_{{user_id}} about
this product: {{item_title}}

Target template: {{explanation}}
```

### Prompt ID: 3-2

```
Input template: Given the following review headline
{{review_headline}}
can you help generate an explanation of user_{{user_id}} for
item_{{item_id}}?

Target template: {{explanation}}
```

### Prompt ID: 3-3

```
Input template: Help user_{{user_id}} generate a {{star_rating}}-star
explanation about this product:
{{item_title}}

Target template: {{explanation}}
```

### Prompt ID: 3-4

```
Input template: Generate an explanation for {{user_desc}} about this
product: {{item_title}}

Target template: {{explanation}}
```

### Prompt ID: 3-5

```
Input template: Based on the following review headline:
{{review_headline}}
Generate {{user_desc}} 's purchase explanation about {{item_title}}

Target template: {{explanation}}
```

### Prompt ID: 3-6

```
Input template: Help {{user_desc}} generate a {{star_rating}}-star
explanation for item_{{item_id}}

Target template: {{explanation}}
```

### Prompt ID: 3-7

```
Input template: Predict the star rating, then use {{feature_word}} as
feature word to generate user_{{user_id}} 's purchase explanation for
item_{{item_id}}

Target template: {{star_rating}}, {{explanation}}
```

### Prompt ID: 3-8

```
Input template: What score will {{user_desc}} rate item_{{item_id}}?
Then give an explanation for the rating score. (1 being lowest and 5
being highest)

Target template: {{star_rating}}, {{explanation}}
```

### Prompt ID: 3-9

```
Input template: Based on the feature word {{feature_word}}, generate
an explanation for user_{{user_id}} about this product:
{{item_title}}

Target template: {{explanation}}
```

### Prompt ID: 3-10

Input template: Given the word {{feature_word}}, can you help generate an explanation for {{user_desc}} about the product: \n {{item_title}}

Target template: {{explanation}}

### Prompt ID: 3-11

Input template: Using the word {{feature_word}}, write a {{star_rating}}-star explanation for user_{{user_id}} about item_{{item_id}}

Target template: {{explanation}}

### Prompt ID: 3-12

Input template: According to the feature word {{feature_word}}, generate a {{star_rating}}-star explanation for {{user_desc}} about item_{{item_id}}

Target template: {{explanation}}

## D.4 Task Family 4: Review Related

### Prompt ID: 4-1

Input template: Write a short sentence to summarize the following product review from user_{{user_id}}:
{{review}}

Target template: {{summary}}

### Prompt ID: 4-2

Input template: Given the following review written by user_{{user_id}}:
{{review}}
Can you predict the associated star rating (1 being lowest and 5 being highest)?

Target template: {{star_rating}}

### Prompt ID: 4-3

Input template: Give a short sentence describing the following product review from {{user_desc}}:
{{review}}

Target template: {{summary}}

### Prompt ID: 4-4

Input template: According to the following review written by {{user_desc}}:
{{review}}
Predict the associated star rating (1 being lowest and 5 being highest)

Target template: {{star_rating}}

## D.5 Task Family 5: Direct Recommendation

### Prompt ID: 5-1

Input template: Will user_{{user_id}} likely to interact with item_{{item_id}}?

Target template: {{answer_choices[label]}} (yes/no)

### Prompt ID: 5-2

Input template: Shall we recommend item_{{item_id}} to {{user_desc}}?

Target template: {{answer_choices[label]}} (yes/no)

### Prompt ID: 5-3

Input template: For {{user_desc}}, do you think it is good to recommend {{item_title}}?

Target template: {{answer_choices[label]}} (yes/no)

### Prompt ID: 5-4

Input template: I would like to recommend some items for user_{{user_id}}. Is the following item a good choice? {{item_title}}

Target template: {{answer_choices[label]}} (yes/no)

### Prompt ID: 5-5

Input template: Which item of the following to recommend for {{user_desc}}?
{{candidate_items}}

Target template: {{target_item}}

### Prompt ID: 5-6

Input template: Choose the best item from the candidates to recommend for {{user_desc}}?
{{candidate_items}}

Target template: {{target_item}}

### Prompt ID: 5-7

Input template: Pick the most suitable item from the following list and recommend to user_{{user_id}}:
{{candidate_items}}

Target template: {{target_item}}

### Prompt ID: 5-8

Input template: We want to make recommendation for user_{{user_id}}. Select the best item from these candidates:
{{candidate_items}}

Target template: {{target_item}}

## D.6 Task Family Z: Zero-Shot Generalization

### Prompt ID: Z-1

Input template: Given the facts about the new product, do you think user {{user_id}} will like or dislike it? title: {{item_title}} brand: {{brand}} price: {{price}}

Target template: {{answer_choices[label]}} (like/dislike) – like (4,5) / dislike (1,2,3)

### Prompt ID: Z-2

Input template: Here are the details about a new product: title: {{item_title}} brand: {{brand}} price: {{price}} What star will {{user_desc}} probably rate the product?
-1 -2 -3 -4 -5

Target template: {{star_rating}}

### Prompt ID: Z-3

Input template: Predict user_{{user_id}} 's preference about the new product (1 being lowest and 5 being highest):
title: {{item_title}} price: {{price}} brand: {{brand}}

Target template: {{star_rating}}

### Prompt ID: Z-4

Input template: Will {{user_desc}} like or dislike the following product? title: {{item_title}} price: {{price}} brand: {{brand}}

Target template:
{{answer_choices[label]}} (like/dislike) – like (4,5) / dislike (1,2,3)

### Prompt ID: Z-5

Input template: Generate a possible explanation for {{user_desc}} 's preference about the following product: title: {{item_title}} brand: {{brand}} price: {{price}}

Target template: {{explanation}}

### Prompt ID: Z-6

Input template: Based on the word {{feature_word}}, help user_{{user_id}} write a {{star_rating}}-star explanation for this new product: title: {{item_title}} price: {{price}} brand: {{brand}}

Target template: {{explanation}}

**Prompt ID: Z-7**

Input template: For the new product {{item_title}}, we would like to know whether {{user_desc}} will love it. If you think the user will love it, please help explain why.

Target template: {{explanation}}

# E FULL LIST OF PERSONALIZED PROMPTS FOR *YELP* DATASET

## E.1 Task Family 1: Rating Prediction

**Prompt ID: 1-1**

Input template: Which star rating will user_{{user_id}} give item_{{item_id}}? (1 being lowest and 5 being highest)

Target template: {{star_rating}}

**Prompt ID: 1-2**

Input template: How will user_{{user_id}} rate this business: {{item_title}}? (1 being lowest and 5 being highest)

Target template: {{star_rating}}

**Prompt ID: 1-3**

Input template: Will user_{{user_id}} give item_{{item_id}} a {{star_rating}}-star rating? (1 being lowest and 5 being highest)

Target template: {{answer_choices[label]}} (yes/no)

**Prompt ID: 1-4**

Input template: Does user_{{user_id}} like or dislike item_{{item_id}}?

Target template: {{answer_choices[label]}} (like/dislike) – like (4,5) / dislike (1,2,3)

**Prompt ID: 1-5**

Input template: Predict the user_{{user_id}} 's preference on item_{{item_id}} ({{item_title}})
-1 \n -2 \n -3 \n -4 \n -5

Target template: {{star_rating}}

**Prompt ID: 1-6**

Input template: What star rating do you think {{user_desc}} will give item_{{item_id}}? (1 being lowest and 5 being highest)

Target template: {{star_rating}}

**Prompt ID: 1-7**

Input template: How will {{user_desc}} rate this business: {{item_title}}? (1 being lowest and 5 being highest)

Target template: {{star_rating}}

**Prompt ID: 1-8**

Input template: Will {{user_desc}} give a {{star_rating}}-star rating for {{item_title}}? (1 being lowest and 5 being highest)

Target template: {{answer_choices[label]}} (yes/no)

**Prompt ID: 1-9**

Input template: Does {{user_desc}} like or dislike {{item_title}}?

Target template: {{answer_choices[label]}} (like/dislike) – like (4,5) / dislike (1,2,3)

**Prompt ID: 1-10**

Input template: Predict {{user_desc}} 's preference towards {{item_title}} (1 being lowest and 5 being highest)

Target template: {{star_rating}}

## E.2 Task Family 2: Sequential Recommendation

**Prompt ID: 2-1**

Input template: Given the following visit history of user_{{user_id}}: {{visit_history}}
predict next possible business to be visited by the user?

Target template: {{next_item}}

**Prompt ID: 2-2**

Input template: I find the visit history list of user_{{user_id}}: {{visit_history}}
I wonder which is the next item to recommend to the user. Can you help me decide?

Target template: {{next_item}}

**Prompt ID: 2-3**

Input template: Here is the visit history list of user_{{user_id}}: {{visit_history}}
try to recommend next item to the user

Target template: {{next_item}}

**Prompt ID: 2-4**

Input template: Given the following visit history of {{user_desc}}: {{visit_history}}
predict next possible business for the user

Target template: {{next_item}}

**Prompt ID: 2-5**

Input template: Based on the visit history of {{user_desc}}: {{visit_history}}
Can you decide the next business likely to be visited by the user?

Target template: {{next_item}}

**Prompt ID: 2-6**

Input template: Here is the visit history of {{user_desc}}: {{visit_history}}
What to recommend next for the user?

Target template: {{next_item}}

**Prompt ID: 2-7**

Input template: Here is the visit history of user_{{user_id}}: {{visit_history}}
Select the next possible business likely to be visited by the user from the following candidates: {{candidate_items}}

Target template: {{next_item}}

**Prompt ID: 2-8**

Input template: Given the following visit history of {{user_desc}}: {{visit_history}}
What to recommend next for the user? Select one from the following items: {{candidate_items}}

Target template: {{next_item}}

**Prompt ID: 2-9**

Input template: Based on the visit history of user_{{user_id}}: {{visit_history}}
Choose the next possible visited business from the following candidates: {{candidate_items}}

Target template: {{next_item}}

**Prompt ID: 2-10**

Input template: I find the visit history list of {{user_desc}}:
{{visit_history}}
I wonder which is the next item to recommend to the user. Try to select one from the following candidates:
{{candidate_items}}

Target template: {{next_item}}

**Prompt ID: 2-11**

Input template: User_{{user_id}} has the following visit history: {{visit_history}}
Does the user likely to visit {{candidate_item}} next?

Target template: {{answer_choices[label]}} (yes/no)

**Prompt ID: 2-12**

Input template: According to {{user_desc}} 's visit history list:
{{visit_history}}
Predict whether the user will visit {{candidate_item}} next?

Target template: {{answer_choices[label]}} (yes/no)

**Prompt ID: 2-13**

Input template: According to the visit history of {{user_desc}}:
{{visit_history}}
Can you recommend the next possible business to the user?

Target template: {{next_item}}

## E.3    Task Family 3: Explanation Generation

**Prompt ID: 3-1**

Input template: Generate an explanation for user_{{user_id}} about this business: {{item_title}}

Target template: {{explanation}}

**Prompt ID: 3-2**

Input template: Help user_{{user_id}} generate a {{star_rating}}-star explanation about this business: {{item_title}}

Target template: {{explanation}}

**Prompt ID: 3-3**

Input template: Generate an explanation for {{user_desc}} about this business: {{item_title}}

Target template: {{explanation}}

**Prompt ID: 3-4**

Input template: Help {{user_desc}} generate a {{star_rating}}-star explanation for item_{{item_id}}

Target template: {{explanation}}

**Prompt ID: 3-5**

Input template: Predict the star rating, then use {{feature_word}} as feature word to generate user_{{user_id}} 's visit explanation for item_{{item_id}}

Target template: {{star_rating}}, {{explanation}}

**Prompt ID: 3-6**

Input template: What score will {{user_desc}} rate item_{{item_id}}? Then give an explanation for the rating score. (1 being lowest and 5 being highest)

Target template: {{star_rating}}, {{explanation}}

**Prompt ID: 3-7**

Input template: Based on the feature word {{feature_word}}, generate an explanation for user_{{user_id}} about this business:
{{item_title}}

Target template: {{explanation}}

**Prompt ID: 3-8**

Input template: Given the word {{feature_word}}, can you help generate an explanation for {{user_desc}} about the business: \n
{{item_title}}

Target template: {{explanation}}

**Prompt ID: 3-9**

Input template: Using the word {{feature_word}}, write a {{star_rating}}-star explanation for user_{{user_id}} about item_{{item_id}}

Target template: {{explanation}}

**Prompt ID: 3-10**

Input template: According to the feature word {{feature_word}}, generate a {{star_rating}}-star explanation for {{user_desc}} about item_{{item_id}}

Target template: {{explanation}}

## E.4    Task Family 4: Review Related

**Prompt ID: 4-1**

Input template: Predict the associated rating score of the review written by user_{{user_id}} (1 being lowest and 5 being highest):
{{review}}

Target template: {{star_rating}}

**Prompt ID: 4-2**

Input template: Given the following review written by user_{{user_id}}:
{{review}}
Can you predict the associated star rating (1 being lowest and 5 being highest)?

Target template: {{star_rating}}

**Prompt ID: 4-3**

Input template: According to the following review written by {{user_desc}}:
{{review}}
Predict the associated star rating (1 being lowest and 5 being highest)

Target template: {{star_rating}}

## E.5    Task Family 5: Direct Recommendation

**Prompt ID: 5-1**

Input template: Will user_{{user_id}} likely to interact with item_{{item_id}}?

Target template: {{answer_choices[label]}} (yes/no)

**Prompt ID: 5-2**

Input template: Shall we recommend item_{{item_id}} to {{user_desc}}?

Target template: {{answer_choices[label]}} (yes/no)

**Prompt ID: 5-3**

Input template: For {{user_desc}}, do you think it is good to recommend {{item_title}}?

Target template: {{answer_choices[label]}} (yes/no)

**Prompt ID: 5-4**

Input template: I would like to recommend some items for user_{{user_id}}. Is the following item a good choice?
{{item_title}}

Target template: {{answer_choices[label]}} (yes/no)

**Prompt ID: 5-5**

```
Input template: Which item of the following to recommend for
{{user_desc}}?
{{candidate_items}}

Target template: {{target_item}}
```

**Prompt ID: 5-6**

```
Input template: Choose the best item from the candidates to recommend
for {{user_desc}}?
{{candidate_items}}

Target template: {{target_item}}
```

**Prompt ID: 5-7**

```
Input template: Pick the most suitable item from the following list
and recommend to user_{{user_id}}:
{{candidate_items}}

Target template: {{target_item}}
```

**Prompt ID: 5-8**

```
Input template: We want to make recommendation for user_{{user_id}}.
Select the best item from these candidates:
{{candidate_items}}

Target template: {{target_item}}
```



\end{document}